\documentclass[twocolumn,prl,aps,floatfix,superscriptaddress,amssymb]{revtex4}
\usepackage{epsfig}
\usepackage{graphics}
\usepackage{amsmath}

\begin{document}

\title{Electronic properties of gated triangular graphene quantum dots: magnetism, correlations and geometrical effects.}

\author{P.~Potasz}
\affiliation{Institute for Microstructural Sciences, National Research Council of Canada
, Ottawa, Canada}
\affiliation{Institute of Physics, Wroclaw University of Technology, Wroclaw, Poland}

\author{A.~D.~G\"u\c{c}l\"u}
\affiliation{Institute for Microstructural Sciences, National Research Council of Canada
, Ottawa, Canada}

\author{A.~W$\rm{\acute{o}}$js}
\affiliation{Institute of Physics, Wroclaw University of Technology, Wroclaw, Poland}

\author{P.~Hawrylak}
\affiliation{Institute for Microstructural Sciences, National Research Council of Canada
, Ottawa, Canada}
\date{\today}

\begin{abstract}
We present a theory of electronic properties of gated triangular graphene
quantum dots with zigzag edges as a function of size and carrier density. We
focus on electronic correlations, spin and geometrical effects using a
combination of atomistic tight-binding, Hartree-Fock and configuration
interaction methods (TB+HF+CI) including long range Coulomb interactions. The
single particle energy spectrum of triangular dots with zigzag edges exhibits a degenerate shell at the Fermi level with a degeneracy
$N_{edge}$ proportional to the edge size. We determine the effect of the
electron-electron interactions on the ground state, the total spin and the excitation spectrum as a function of a shell filling and the degeneracy of the shell using
TB+HF+CI for $N_{edge} < 12$ and approximate CI method for $N_{edge}\geq
12$. For a half-filled neutral shell we find spin polarized ground state for
structures up to $N=500$ atoms in agreement with previous {\it ab initio} and
mean-field calculations, and in agreement with Lieb's theorem for a Hubbard
model on a bipartite lattice. Adding a single electron leads to the complete spin
depolarization for $N_{edge} \leq 9$. For larger structures, the spin
depolarization is shown to occur at different filling factors. Away from
half-fillings excess electrons(holes) are shown to form Wigner-like spin
polarized triangular molecules corresponding to large gaps in the excitation
spectrum. The validity of conclusions is assessed by a comparison of results
obtained from different levels of approximations. While for the charge neutral
system all methods give qualitatively similar results, away from the charge
neutrality an inclusion of all Coulomb scattering terms  is necessary to produce
results presented here. 
\end{abstract}
\maketitle

\section{I. Introduction}

Graphene is an atomically thick layer of carbon atoms arranged in a honeycomb
lattice.\cite{Novoselov+Geim+04,Novoselov+Geim+05,Zhang+Tan+05,Son+PRL+06,Potemski+deHeer+06,
Geim+Novoselov+07,Rycerz+Tworzydlo+07,Xia+Mueller+09,Mueller+Xia+10,Neto+Guinea+09}
Due to its unique electronic properties and promising potential for
applications, there is a growing research interest in graphene based
nanostructures.\cite{Neto+Guinea+09,Abergel+10,Rozhkov+11} Attempts at
fabricating graphene nanostructures with well defined shape and edge type have
been reported starting from the graphene layer and using top-down techniques,
\cite{Li+08,Ponomarenko+08,Ci+08,You+08,Schnez+09,Ritter+09,Jia+09,Campos+09,Neubeck+10,Biro+10,CruzSilva+10,Yang+10,Krauss+10}
bottom-up techniques \cite{Zhi+08,Treier+10,Mueller+10,Morita+11,Lu+11}
starting from carbon based molecules, as well as starting from graphane and
removing hydrogen atoms using AFM tips.\cite{Singh+09,Tozzini+10,Xiang+09,Schmidt+10} 

The work on graphene nanostructures is motivated by the expectation that
finite size effects significantly modify electronic properties of graphene.
As a result of size quantization, an energy gap opens up,  making graphene a
semiconductor with a gap tunable from THz to UV.  The energy gap can be tuned
by changing not only the size but also the shape and the type of edge, allowing us to control the material's optical properties.\cite{Yamamoto+06,Zhang+08,Guclu+10} Two types of edges in graphene  are of
particular interest due to their stability: armchair and zigzag. For zigzag
edges, edge states in the vicinity of the Fermi energy appear. This is related
to breaking the sublattice symmetry between the two types of atoms in the unit cell of the graphene honeycomb lattice. The presence of edge states was predicted
theoretically
\cite{NFD+96,Fujita+96,Son+06,Son+PRL+06,Ezawa+06,Yamamoto+06,Ezawa+07,FRP+07,AHM+08,Wang+Yazyev+09,Potasz+10}
and confirmed experimentally.\cite{Niimi+Matsui+05,Kobayashi+Fukui+05,Tao+Jiao+11} These
edge states form a degenerate band in graphene ribbons
\cite{NFD+96,Fujita+96,Son+06,Son+PRL+06,Ezawa+06} or can collapse to a
degenerate shell in graphene quantum dots.\cite{Yamamoto+06,Ezawa+07,FRP+07,AHM+08,Wang+Meng+08,Wang+Yazyev+09,Guclu+09,Potasz+10} It
was previously shown that the degeneracy is equal to the difference between
the number of atoms corresponding to two sublattices in the bipartite lattice.\cite{Ezawa+07,FRP+07,Wang+Meng+08,Potasz+10} In particular, the geometry that maximizes the imbalance between the two sublattices is a zigzag edge
triangle where the degeneracy of the zero-energy shell is proportional to
the number of atoms on the one edge.\cite{Potasz+10} This presents a unique
opportunity to design a quantum system with a macroscopic degeneracy, analogously
to the two-dimensional electron gas in a strong magnetic field. 

While fabricating and measuring triangular graphene quantum dots with well defined edges \cite{Campos+09,Zhi+08,Lu+11,Morita+11} remains a challenge, the theory of triangular graphene quantum dots (TGQD) with zigzag edges was developed  by several groups.\cite{Yamamoto+06,Guclu+10,Ezawa+07,FRP+07,AHM+08,Wang+Meng+08,Ezawa+08,Philpott+08,HMA+08,Guclu+09,Potasz+10,Kosimov+10,Ezawa+10,Sahin+10,Ezawa+E10,Voznyy+11,Morita+11,Romanovsky+11,Xi+09,Kinza+10,Zarenia+11,Dai+12} 
In particular, the macroscopically degenerate zero-energy band and the
corresponding wavefunctions were explicitely constructed.\cite{Potasz+10} For a half-filled shell, TGQDs were studied by Ezawa using the Heisenberg Hamiltonian, \cite{Ezawa+07} by Fernandez-Rossier and Palacios \cite{FRP+07} using the mean-field Hubbard model, by Wang, Meng, and Kaxiras \cite{Wang+Meng+08} using density functional theory (DFT); and G\"u\c{c}l\"u {\it et al.}  \cite{Guclu+09} using exact diagonalization techniques. It was shown  that the ground state is fully spin polarized, with a finite magnetic moment proportional to the shell degeneracy. This finding is in agreement with Lieb's theorem on magnetism of the Hubbard model for bipartite lattice systems.\cite{Lieb+89} 

The effect of defects and disorder was also investigated.\cite{Potasz+10,Ezawa+E10,Voznyy+11} In particular, Voznyy {\it et al.}
\cite{Voznyy+11} have shown by using {\it ab initio} methods that
hydrogen-passivation stabilizes zigzag edges in TGQD over the
pentagon-heptagon reconstructed edges.\cite{Voznyy+11} It was  also proved
that the zero-energy shell survives when TGQD is deformed to trapezoidal shape.\cite{Potasz+10} Ezawa studied the stability of magnetization against
disorder. He considered three types of randomness: in a hopping integral, a
site energy and lattice defects.\cite{Ezawa+E10} He proved that the magnetism
is still governed by Lieb's theorem but the number of degenerate states
changed by the number of lattice defects. Some of us have shown in
Ref. \onlinecite{Guclu+09} by use of methods beyond mean-field approximations,
that the magnetization is unstable with respect to additional charge, leading to a
complete spin depolarization. The spin depolarization was shown  to
significantly influence transport properties, blocking current through the
graphene quantum dot due to the spin blockade.\cite{Guclu+09} It was also
shown that by changing the population of the degenerate shell using a gate, one
can simultaneously control magnetic and optical properties, determined by
strong electron-electron and excitonic interactions.\cite{Guclu+10}

In this work we use improved configuration-interaction (CI) tools to extend
our previous results \cite{Guclu+09} regarding the role of electron-electron
interactions, magnetism and correlations in TGQDs to larger structures. We
investigate the electronic properties as a function of size and filling factor
of the degenerate shell by using a combination of tight-binding (TB),
Hartree-Fock (HF) and configuration
interaction methods (TB+HF+CI). Our many-body Hamiltonian includes, in addition to
the on-site interaction term, all scattering and exchange terms within
next-nearest neighbors, and all direct interaction terms in the two-body
Coulomb matrix elements. Using full CI combined with the TB+HF method
we demonstrate that the ground state for the charge neutral system has
maximally polarized edge states for structures consisting of up to 200 atoms
with the number of degenerate edge states $N_{edge} \leq 9$.  By analyzing a spin-flip excitation spectrum of the spin-polarized ground state, we verify the spin-polarized ground state for up to 500 atoms or $N_{edge}=20$. These results for a system with long ranged Coulomb interaction  appear
to be consistent with Lieb's theorem for the Hubbard model. Using TB+HF+full
CI method for TGQD charged with an additional electron and a size of up to $N=200$
atoms it is shown that a complete spin depolarization predicted earlier by
some of us \cite{Guclu+09} exists only up to a critical size. The critical
size is established by studying the stability of a charged spin-polarized shell
to spin-flip excitations. It is shown that for sizes up to the critical size the
spin wave and minority spin electron form a bound state, a trion, signaling
the tendency to the depolarization. For sizes exceeding the critical size the
spin waves are unbound and the spin-polarized state is the ground state up to the
sizes studied ($N\approx 500$ atoms). For TGQD structures above the critical size,
depolarization effects away from the half-filling are observed. Results of TB+HF+CI
calculations allow us to extract the excitation gap as a function of a shell
filling. It is found that the largest gaps correspond to the half-filled spin-polarized shell and special filling fractions. At these special filling
fractions, we predict a formation of Wigner-like spin polarized molecules,
related to long range Coulomb interactions and a triangular geometry of
graphene quantum dot. Finally, we compare results obtained at different levels
of approximations. We show that, for the charge neutral system, the Hubbard,
extended Hubbard, and fully interacting models are in good qualitative
agreement. On the other hand, away from the half-filling, only a fully interacting
model is able to correctly capture the effect of correlations.   
  
The paper is organized as follows. In Sec. II, we describe our
model. Section III contains analysis of the ground state spin and correlations
as a function of size and filling factor of the degenerate shell. In
Sec. IV, we compare results obtained within different levels of
approximations. In Sec. V, we summarize our results.

\section{II. Model of a graphene triangular quantum dot }
\begin{figure}
\epsfig{file=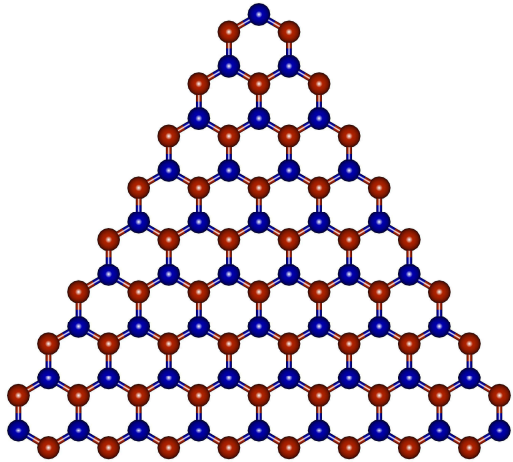,width=2.8in}
\caption{(Color online) Triangular graphene quantum dot with zigzag
  edges. There are eight edge atoms (with two bonds) on one edge. Red
  (light gray) and blue (dark gray) colors distinguish between two sublattices
  in the honeycomb graphene lattice. Structure consists of a total of $N=97$ atoms}
\label{fig:Fig1}
\end{figure}

Graphene is a two-dimensional honeycomb crystal formed by carbon atoms on  two
interpenetrating hexagonal sublattices. The unit cell thus contains two carbon
atoms.  The distance between nearest-neighbor atoms is $a=1.42$ $\rm{\AA}$. By
using vectors ${\bf{R}}=n{\bf{a}}_{1}+m{\bf{a}}_{2}$ with $n,m$ integers and
primitive unit vectors defined as ${\bf{a}}_{1,2}=a/2(\pm \sqrt{3},3)$, one
can obtain the positions of all the atoms in the structure. By cutting the graphene lattice in three zigzag directions, an equilateral triangle can be
obtained, as shown in Fig. \ref{fig:Fig1}. Such a system has a broken sublattice
symmetry with two properties: (i) all edge atoms (with only two bonds) belong
to the same sublattice, (ii) the difference between the number of atoms
belonging to each sublattice is proportional to the number of atoms on one of
the three edges. 

\begin{figure}
\epsfig{file=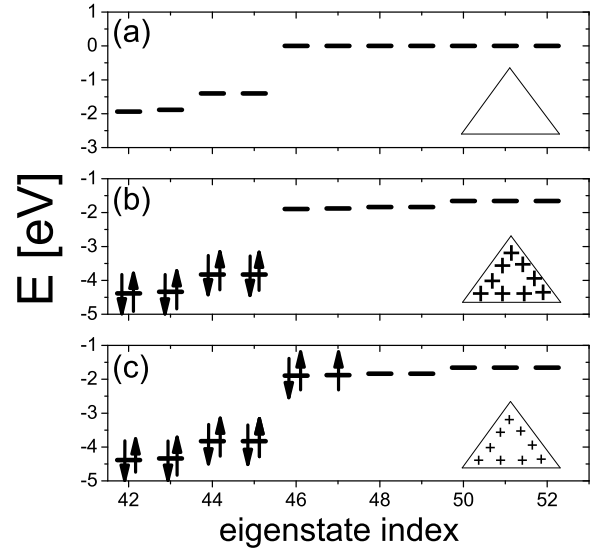,width=3.4in}
\caption{(a) Single-particle nearest-neighbor tight-binding (TB) energy
  levels. The zero-energy shell on the Fermi level is perfectly
  degenerate. (b) Positively charged system with an empty degenerate band after
  self-consistent Hartee-Fock (HF) mean-field calculations described by a
  single Slater determinant (TB+HF model). (c) Occupation of empty degenerate
  HF quasi-orbitals by electrons. The inset pictures schematically show the
  excess charge corresponding to each of the three model systems. The ground state and
  the total spin of the system of interacting electrons can be calculated by
  using the configuration interaction (CI) method, described in Sec. II
  (TB+HF+CI). The charge neutrality corresponds to a half-filled degenerate band
  (not shown).}
\label{fig:fig2}
\end{figure}

Each carbon atom has four valence electrons. Three of them, on $s$, $p_x$, and
$p_y$ orbitals, form $sp^2$ bonds with the three nearest in-plane neighbors.
They are strongly bound and responsible for mechanical properties of
graphene. The remaining fourth valence electron on  each carbon atom  $p_z$
orbital, perpendicular to the plane of graphene, is weakly bound and
determines electronic properties of the system. Single-particle properties of
graphene  can be described by using one orbital tight-binding (TB) Hamiltonian.\cite{Wallace+47} We have previously shown that, within the TB model in the
nearest-neighbors approximation,  TGQDs with zigzag edges exhibit an energy
gap, with a degenerate shell at the Fermi (zero) energy, with a degeneracy
proportional to the length of an edge.\cite{Potasz+10} An example of TB
energy levels  for a structure consisting of 97 atoms with $N_{edge}=7$
degenerate states is shown in Fig. \ref{fig:fig2}(a). Our goal is to study the
role of electron-electron interactions for electrons occupying this degenerate
shell. Solving the full many-body problem even for such a small structure with
97 atoms is
not possible at present. However, due to the energy gap separating the valence
band and degenerate states, the valence electrons that do not occupy the
degenerate band can be treated in a mean-field approximation.  The remaining
electrons occupying the degenerate shell must, however, be treated using a configuration-interaction method (CI). Therefore, we use a TB+HF+CI approach
that allows us to treat the electronic correlations for electrons in the
degenerate shell and their interaction with valence electrons at the
mean-field level.

We start from the full many-body Hamiltonian for interacting electrons on the
$p_z$ orbitals of graphene. It can
be written as
\begin{eqnarray}
H= \sum_{i,l,\sigma}\tau_{il\sigma}c^\dagger_{i\sigma}c_{l\sigma}
    +\frac{1}{2}\sum_{\substack{i,j,k,l,\\\sigma \sigma'}}\langle ij\vert V \vert kl\rangle 
     c^\dagger_{i\sigma}c^\dagger_{j\sigma'}c_{k\sigma'}c_{l\sigma},
\label{fullH}
\end{eqnarray}
where the operator $c^\dagger_{i\sigma}$ creates an electron on the $i$-th $p_z$
orbital with spin $\sigma$, $\tau_{il\sigma}$ is
a hopping integral that describes the probability of scattering of electron
on the $l$-th $p_z$ orbital $\phi_{l}$ to the $i$-th $p_z$ orbital $\phi_{i}$. The second
term describes two-body Coulomb interactions between $p_z$ electrons. Note that at this stage, the unknown hopping terms $\tau_{il\sigma}$ do not
include the effect of electron-electron interactions of $p_z$ electrons.

\subsection{A. Mean-Field approximation for infinite graphene sheet}
Let us first write the Hamiltonian for graphene, given by Eq. (\ref{fullH}), in the mean-field Hartree-Fock (HF) approximation  as
\begin{eqnarray}
\nonumber
H_{MF}^{o}&=& \sum_{i,l,\sigma}\tau_{il\sigma}c^\dagger_{i\sigma}c_{l\sigma} 
+ \sum_{i,l,\sigma} \sum_{j,k,\sigma'}\rho_{jk\sigma'}^{o}
(\langle ij\vert V \vert kl\rangle \\
&-&\langle ij\vert V \vert lk\rangle\delta_{\sigma,\sigma'})        c^\dagger_{i\sigma}c_{l\sigma}
=\sum_{i,l,\sigma}t_{il\sigma}c^\dagger_{i\sigma}c_{l\sigma}. 
\label{HMF}
\end{eqnarray}
This is effectively a one-body TB Hamiltonian for a graphene layer \cite{Wallace+47} with density
matrix elements $\rho_{jk\sigma'}^{o}=\langle c^\dagger_{j\sigma '}c_{k\sigma '}\rangle$ calculated with respect to the fully occupied valence band. The values of
$\rho^{o}_{jk\sigma'}$ are evaluated in Appendix A and their role becomes clear
in the next subsection. $t_{il\sigma}$ are experimentally estimated hopping
integrals. 

\subsection{B. Mean-Field approximation for graphene quantum dots}
We now derive a mean-field Hamiltonian for electrons in graphene quantum
dots (GQD). First, we apply mean-field approximation to the Hamiltonian given
by Eq. (\ref{fullH}) for electrons in GQD, with a result written as
\begin{eqnarray}
\nonumber
H_{MF}^{GQD}&=& \sum_{i,l,\sigma}\tau_{il\sigma}c^\dagger_{i\sigma}c_{l\sigma} 
+ \sum_{i,l,\sigma} \sum_{j,k,\sigma'}\rho_{jk\sigma'}
(\langle ij\vert V \vert kl\rangle \\
&-&\langle ij\vert V \vert lk\rangle\delta_{\sigma,\sigma'})        c^\dagger_{i\sigma}c_{l\sigma}, 
\label{HDOT}
\end{eqnarray}
with density matrix $\rho_{jk\sigma'}$ for GQD. By combining Eq. (\ref{HMF}) and Eq. (\ref{HDOT}) we get  
\begin{eqnarray}
\nonumber
&H_{MF}^{GQD}&=H_{MF}^{GQD}-H_{MF}^{o}+H_{MF}^{o} \\
\nonumber
&=& \sum_{i,l,\sigma}t_{il\sigma}c^\dagger_{i\sigma}c_{l\sigma} 
     + \sum_{i,l,\sigma} \sum_{j,k,\sigma'}(\rho_{jk\sigma'}-\rho^{o}_{jk\sigma'}) 
(\langle ij\vert V \vert kl\rangle \\
&-&\langle ij\vert V \vert lk\rangle\delta_{\sigma,\sigma'})c^\dagger_{i\sigma}c_{l\sigma}. 
\label{hfhamilton}
\end{eqnarray}
Here the subtracted component in the second term corresponds to mean-field
interactions included in effective $t_{il\sigma}$ hopping integrals, described
by the graphene density matrix $\rho^{o}_{jk\sigma'}$. For the TGQDs, the density matrix elements $\rho_{jk\sigma'}$ are calculated with respect to the many-body
ground state of $N_{ref}=N_{site}-N_{edge}$ electrons, where $N_{site}$ is the
number of atoms. Since the valence band and the degenerate shell are separated
by an energy gap, the closed shell system of $N_{ref}$ interacting electrons
is expected to be well described in a mean-field approximation, using a single
Slater determinant. This corresponds to a charged system with $N_{edge}$
positive charges, as schematically shown in Fig. \ref{fig:fig2}(b). The
Hamiltonian given by Eq. (\ref{hfhamilton}) has to be solved self-consistently
to obtain Hartree-Fock quasi particle orbitals. In numerical calculations, in
addition to the on-site interaction term, all scattering and exchange terms
within next-nearest neighbors and all direct interaction terms are included
in the two-body Coulomb matrix elements $\langle ij|V|kl\rangle$ computed
using Slater $p_z$ orbitals.\cite{Ransil+60} The few largest Coulomb matrix
elements are given in Ref. \onlinecite{Potasz+ring+10}. The value of the
effective dielectric constant $\kappa$ depends on the substrate and is set to
$\kappa=6$ in what follows.\cite{Reich+02} A method of calculating values of
$\rho^{o}_{jk\sigma'}$ for graphene is shown in Appendix A. Values of
hopping integrals $t_{il\sigma}$ are taken from the experimental data
\cite{Bostwick+07} or {\it ab initio} calculation.\cite{Reich+02} We use
$t=-2.5$ eV for nearest-neighbors and $t'=-0.1$ eV for next-nearest-neighbors
\cite{DCN+07} hopping matrix elements. The HF results were also compared with
the results of {\it ab initio} calculations.\cite{Guclu+09,Voznyy+11}

We now discuss mean-field results for the charge neutral system.  In the
vicinity of the center of a sufficiently large dot a charge distribution
around a given site is identical to that of an infinite system. The density
matrices for graphene $\rho^{o}_{jk\sigma'}$ and for GQD $\rho_{jk\sigma'}$ are equal. A second term in
Eq. (\ref{hfhamilton}) vanishes, leaving only a hopping integral
$t_{il\sigma}$. On the other hand, close to the edges, a density matrix for
the GQD differs in comparison to its graphene counterpart. After
diagonalizing the HF Hamiltonian given by Eq. (\ref{hfhamilton}) one obtains
eigenvalues and eigenvectors that involve the geometrical properties of the
system, shown in Fig. \ref{fig:fig2}(b). A slight removal of the degeneracy of
middle edge states and three corner states with a bit higher energies are
observed, with electronic densities shown in Ref. \onlinecite{Guclu+09}. 

\subsection{C. Configuration Interaction method}
After the self-consistent procedure we get new orbitals for quasi-particles with a fully occupied valence band and a completely empty degenerate shell. We start
filling these degenerate states by adding extra electrons one by one, schematically shown in Fig. \ref{fig:fig2}(c). Next, we
solve the many-body Hamiltonian corresponding to the added electrons, given by
\begin{eqnarray}
\nonumber
&H_{MB}&=\sum_{s,\sigma}\epsilon_{s}a^\dagger_{s\sigma}a_{s\sigma}
\\&+&\frac{1}{2}\sum_{\substack{s,p,d,f,\\\sigma,\sigma'}} \langle sp\mid V\mid df\rangle
a^\dagger_{s\sigma}a^\dagger_{p\sigma'}a_{d\sigma'}a_{f\sigma},       
\label{MBody}
\end{eqnarray}
where the first term describes the energies of the Hartree-Fock orbitals and the second term
describes an interaction between quasi particles occupying degenerate HF states
denoted by $s,p,d,f$ indices. The two-body quasi particle scattering matrix
elements $\langle sp\mid V\mid df\rangle$  are calculated from the two-body
localized on-site Coulomb matrix elements $\langle ij\mid V\mid kl\rangle$.

In our calculations, we neglect scattering from/to the states from a fully
occupied valence band. Moreover, because of the large energy gap between the
shell and the conduction band, we can neglect scatterings to the higher energy
states. A validity of these approximations is assessed in
Ref. \cite{Potasz+ring+10}. These approximations allow us to treat the degenerate shell as an
independent system that significantly reduces the dimension of the Hilbert
space. The basis is constructed from vectors corresponding to all possible
many-body configurations of electrons distributed within the degenerate
shell. For a given number of electrons $N_{el}$, the Hamiltonian given by
Eq. (\ref{MBody}) is diagonalized in each subspace with total
 $S_{z}$. 

\subsection{D. Effect of gate charge}
In our model, we start from the system with an empty shell that corresponds
to the charged system. As in our previous work, Ref. \onlinecite{Guclu+09},
electrons from the shell are transferred to the metallic gate. The Hamiltonian for
$N_{ref}$ electrons in the presence of a gate in the mean-field Hartree-Fock
approximation was written as
\begin{eqnarray}
\nonumber
&H_{MF}&= \sum_{i,l,\sigma}t_{il\sigma}c^\dagger_{i\sigma}c_{l\sigma} 
     + \sum_{i,l,\sigma} \sum_{j,k,\sigma'}(\rho_{jk\sigma'}-\rho^{o}_{jk\sigma'}) 
(\langle ij\vert V \vert kl\rangle \\
&-&\langle ij\vert V \vert lk\rangle\delta_{\sigma,\sigma'})
           c^\dagger_{i\sigma}c_{l\sigma} 
+ \sum_{i,\sigma}v^{g}_{ii}(q_{ind})c^\dagger_{i\sigma}c_{i\sigma}, 
\label{hfgate}
\end{eqnarray}
with an electrostatic potential $v_{ii}^{g}$ related to $N_{edge}$ electrons in a gate defined as
\begin{eqnarray}
v_{ii}^{g}(q_{ind})&=& \sum^{N_{site}}_{j=1}\frac{-q_{ind}/N_{site}}
{\kappa\sqrt{(x_i-x_j)^2+(y_i-y_j)^2+d^2_{gate}}}
\label{Hgate}
\end{eqnarray} 
with $q_{ind}=-N_{edge}$ charge smeared out at positions $(x_i,y_i)$ at a
distance $d_{gate}$ from the quantum dot. Next, we derived the many-body
Hamiltonian with an inclusion of the effect of gate, written as 
\begin{eqnarray}
\nonumber
H=\sum_{p,\sigma}\epsilon_{p}a^\dagger_{p\sigma}a_{p\sigma}
    +\frac{1}{2}\sum_{\substack{p,q,r,s,\\\sigma,\sigma'}}\langle pq\vert V\vert rs\rangle 
     a^\dagger_{p\sigma}a^\dagger_{q\sigma'}a_{r\sigma'}a_{s\sigma} \\
+\sum_{p,q,\sigma} \langle p\vert v^{g}(N_{add})\vert q\rangle
     a^\dagger_{p\sigma}a_{q\sigma}+ 2\sum_{p'}\langle p'\vert v^{g}(N_{add})\vert p'\rangle,
\label{CIhamilton}
\end{eqnarray}
where the indices without the prime sign $(p,q,r,s)$ run over $N_{edge}$
degenerate states, while the index with the prime sign $p'$ runs over
$N_{ref}/2$ valence states (below the degenerate shell). A third term in
Eq. (\ref{CIhamilton}) corresponds to scattering from state $q$ to state $p$ in
a degenerate shell as a result of interactions with electrons in a gate. The
fourth term is a constant and just shifts the entire spectrum by a constant
energy. 

\section{III. Magnetism and Correlation effects}

\subsection{Electronic properties as a function of the filling factor}
\begin{figure}
\epsfig{file=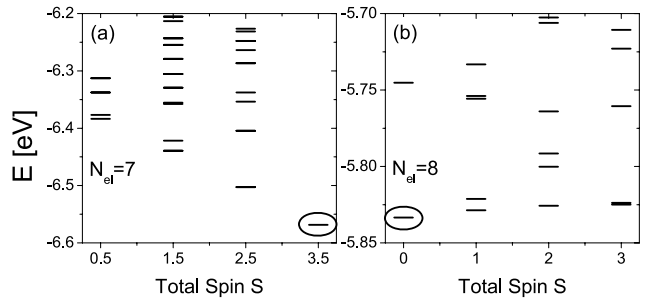,width=3.4in}
\caption{The low-energy spectra for the different total spin $S$ for (a)
  $N_{el}=7$ electrons and (b) $N_{el}=8$ electrons. For $N_{el}=7$ electrons
  the ground state corresponding to $S=3.5$, indicated by a circle, is well
  separated from excited states with different total spin $S$. For $N_{el}=8$
  electrons the ground state corresponding to $S=0$, indicated by a circle, is
  almost degenerate with excited states with different total spin $S$.} 
\label{fig:fig3}
\end{figure}   
We, first, concentrate on a TGQD consisting of $N=97$ atoms, which is the
largest system previously studied in our earlier work in
Ref. \cite{Guclu+09}. It has $N_{edge}=7$ zero-energy degenerate states
obtained from TB calculations, shown in Fig. \ref{fig:fig2}(a). After
self-consistent HF calculations neglecting the gate charge (the effect of the
gate will be discussed later), we obtain new quasi particle orbitals, shown
in Fig. \ref{fig:fig2}(b). The degeneracy is slightly removed. We fill these
degenerate levels by additional electrons and calculate two-body scattering
matrix elements. For a given number of quasi particles, the many-body
Hamiltonian, Eq. (\ref{MBody}), is diagonalized in a basis of configurations of
electrons distributed within the shell, as explained in Sec. II.  In
Fig. \ref{fig:fig3}, we analyze the dependence of the low energy spectra on
the total spin $S$ for [Fig. \ref{fig:fig3}(a)] the charge neutral system, $N_{el}=7$ electrons,
and [Fig. \ref{fig:fig3}(b)] one added electron, i.e., $N_{el}=8$ electrons. We see that for the
charge neutral TGQD with $N_{el}=7$ electrons the ground state of the system
is maximally spin polarized, with $S=3.5$, indicated by a circle. There is
only one possible configuration of all electrons with parallel spins that
corresponds to exactly one electron per one degenerate state. The energy
of this configuration is well separated from other states with lower  total
spin $S$, which require at least one flipped spin among seven initially spin-polarized electrons. An addition of one extra electron to the system with
$N_{el}=7$ spin polarized electrons induces correlations as seen in
Fig. \ref{fig:fig3}(b), where the cost of flipping one spin is very
small. Moreover, for $N_{el}=8$, the ground state is completely depolarized
with $S=0$, indicated by a circle,  but this ground state is almost degenerate
with states corresponding to the different total spin. 

\begin{figure}
\epsfig{file=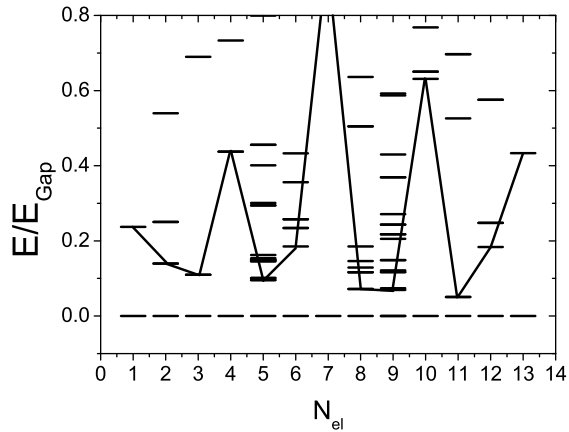,width=3.4in}
\caption{The low-energy spectra of the many-body states as a function of the
  number of electrons occupying the degenerate shell for the system with
  $N_{edge}=7$ degenerate states. The energies are renormalized by the energy
  gap corresponding to the half-filled shell, $N_{el}=7$ electrons. A large
  density of states around $N_{edge}+1$ electrons is a signature of the
  correlation effects. The solid line shows the evolution of the energy gap as
  a function of shell filling.} 
\label{fig:fig4}
\end{figure}   
The calculated many-body energy levels, including all spin states for different
numbers of electrons (shell filling), are shown in Fig. \ref{fig:fig4}. For each electron
number, $N_{el}$, energies are measured from the ground-state energy and
scaled by the energy gap of the half-filled shell, corresponding to $N_{el}=7$
electrons in this case. The solid line shows the evolution of the energy gap
as a function of shell filling. The energy gaps for a neutral system ,
$N_{el}=7$ , as well as for $N_{el}=7-3=4$ and $N_{el}=7+3=10$ are found to be
significantly larger in comparison to the energy gaps for other electron
numbers. In addition, close to the half-filled degenerate shell, the reduction
of the energy gap is accompanied by an increase of low energy density of
states. This is a signature of correlation effects, showing that they can
play an important role at different filling factors.

\begin{figure}
\epsfig{file=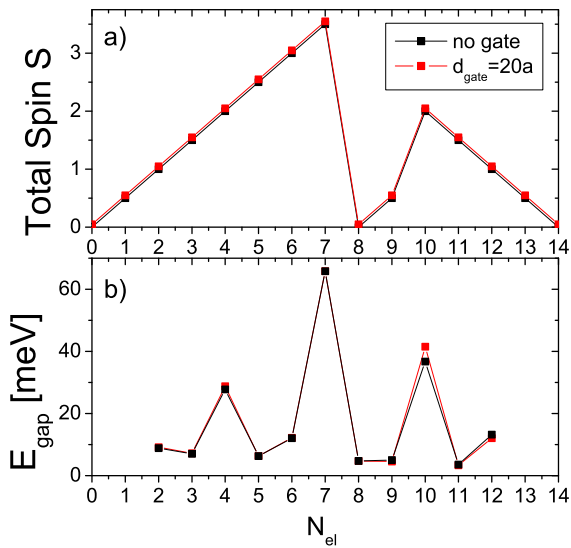,width=3.4in}
\caption{(Color online) (a) The total spin as a function of number of electrons occupying the degenerate shell and (b) corresponding the energy excitation gaps, with and without a gate, red (light gray) and black (dark gray) lines, respectively. Due to a presence of correlation effects for some
  fillings, the magnitude of the energy gap is significantly reduced.} 
\label{fig:fig5}
\end{figure}   
We now extract the total spin and energy gap for each electron number. Figures
\ref{fig:fig5}(a) and (b) show the phase diagram, the total spin $S$ and an excitation gap as a function of the number of electrons occupying the
degenerate shell. The system reveals maximal spin polarization for almost all
fillings, with exceptions for $N_{el}=8,9$ electrons. However, the energy
gaps are found to strongly oscillate as a function of shell filling as a
result of a combined effect of correlations and system's geometry. We observe
a competition between fully spin polarized system that maximizes exchange
energy and fully unpolarized system that maximizes the correlation
energy. Only close to the charge neutrality, for $N_{el}=8$ and $N_{el}=9$
electrons, are the correlations sufficiently strong to overcome the large cost
of the exchange energy related to flipping spin. The excitation gap is
significantly reduced and exhibits large  density of states at low energies,
as shown in Fig. \ref{fig:fig3}.

\begin{figure}
\epsfig{file=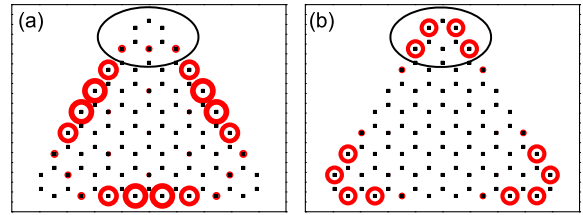,width=3.4in}
\caption{(Color online) The spin densities of the ground state for (a)
  $N_{el}=4$ electrons and (b) $N_{el}=10$ electrons that correspond to
  subtracting/adding three electrons from/to the charge neutral system. The radius
  of circles is proportional to a value of spin density on a given atom. A
  long range Coulomb interaction repels (a) holes and (b) electrons to
  three corners, forming a spin-polarized Wigner-like molecule.} 
\label{fig:fig6}
\end{figure}   
Away from half-filling, we observe larger  excitation gaps for $N_{el}=4$ and
$N_{el}=10$ electrons. These fillings correspond to subtracting/adding three electrons from/to the charge-neutral system with $N_{el}=7$ electrons. In
Fig. \ref{fig:fig6} we show the corresponding spin densities. Here, long range
interactions dominate the physics and three spin polarized [Fig. \ref{fig:fig6}(a)] holes
($N_{el}=7-3$ electrons) and [Fig. \ref{fig:fig6}(b)] electrons ($N_{el}=7+3$ electrons) maximize
their relative distance by occupying three consecutive corners. Electron spin
density is localized in each corner while holes correspond to missing spin
density localized in each corner. We also note that this is not observed for
$N_{el}=3$ electrons filling the degenerate shell (not shown here). The
energies of HF orbitals of corner states correspond to three higher energy
levels [see Fig. \ref{fig:fig12}(c)], with electronic densities shown in
Ref. \cite{Guclu+09}. Thus, $N_{el}=3$ electrons occupy lower-energy
degenerate levels corresponding to sides instead of corners. On the other
hand, when $N_{el}=7$ electrons are added to the shell, self-energies of extra
electrons renormalize the energies of HF orbitals. The degenerate shell is
again almost perfectly flat similarly to levels obtained within the TB
model. A kinetic energy does not play a role allowing a formation of a spin-polarized Wigner-like molecule, resulting from a long-range interactions and a
triangular geometry. We note that Wigner molecules were previously discussed in circular graphene quantum dots with zigzag edges described in the  effective mass approximation.\cite{Wunsch+2008,Romanovsky+2009} The rotational symmetry of quantum dot allowed for the construction of an approximate correlated ground state corresponding to either a Wigner-crystal or Laughlin-like state.\cite{Wunsch+2008} Later, a variational rotating-electron-molecule (VREM) wave function was used.\cite{Romanovsky+2009} Unfortunately, due to a lack of an analytical form of a correlated wave function with a triangular symmetry, it is not possible to do it here.

\begin{figure}
\epsfig{file=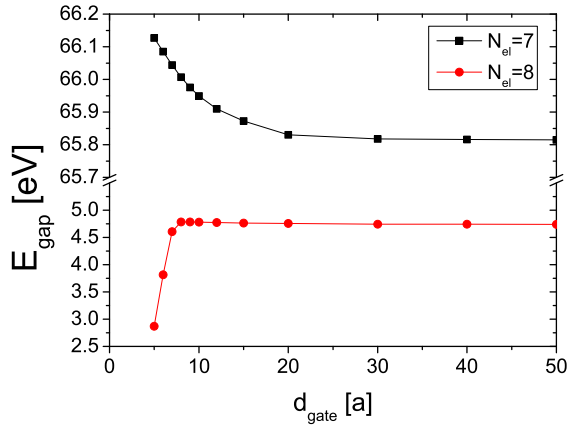,width=3.4in}
\caption{The energy gaps around the charge neutrality for a system with
  $N_{edge}=7$ degenerate states as a function of a gate distance. The energy
  gap for the charge-neutral system, $N_{el}=7$, changes by less than
  $1\%$. For the charged system, $N_{el}=8$, we observe changes in the energy
  gap for a gate distance in a range $5a \leq d_{gate}\leq 10a$ but still not
  affecting the spin depolarization.}
\label{fig:fig7}
\end{figure}   
Figure \ref{fig:fig5} also shows the effect of the presence of a gate at a
distance $d_{gate}=20a$, where $a=1.42$ is a nearest-neighbor's inter-atomic
distance. Clearly, the effect of a gate is very weak, just slightly changing
energy gaps. In Fig. \ref{fig:fig7}, energy gaps as a function of a gate
distance for the charge-neutral $N_{el}=7$ and charged $N_{el}=8$ system for
our tested system with $N_{edge}=7$ degenerate states are shown. There are no
effects for a gate distance $d_{gate}\geq 20a$. When a gate distance is
comparable to graphene-substrate separation, $d_{gate}\sim 5a$, the energy gap for $N_{el}=7$ increases while the energy gap for $N_{el}=8$ decreases. The drop for $N_{el}=8$ is not sufficiently strong to change an observed effect of the spin depolarization. According to the above analysis, we next present results for a Hamiltonian with a gate at infinity.

\subsection{Electronic properties as a function of the size}
In a previous section, we have analyzed in detail the electronic properties of
a particular TGQD with $N=97$ atoms  as a function of the filling factor
$\nu=N_{el}/N_{edge}$, i.e., the number of electrons per number of degenerate
levels. In this section we address the important question of whether one can
predict the electronic properties of a TGQD as a function of size. 

\begin{figure}
\epsfig{file=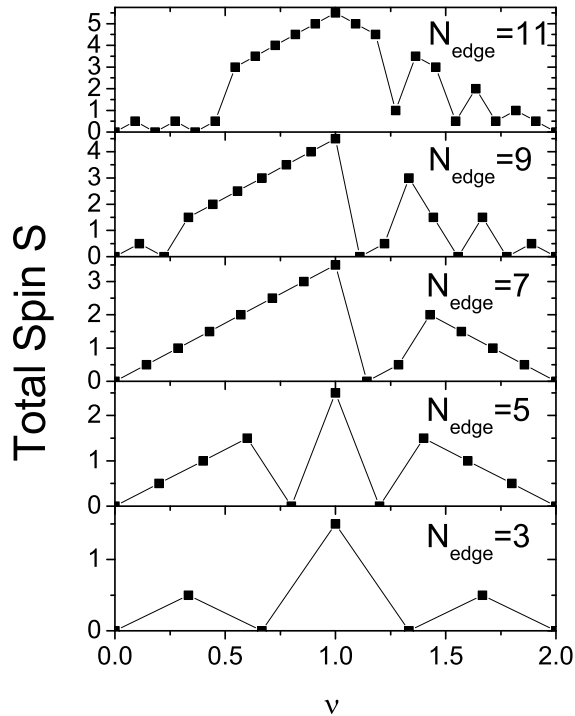,width=3.4in}
\caption{ Spin phase diagrams as a function of filling factor $\nu=N_{el}/N_{edge}$ for different size triangles characterized by the number of the degenerate edge states $N_{edge}$. Half-filled shell $\nu=1$ is always maximally spin polarized. The complete spin depolarization occurs for one added electron to the charge-neutral system for $N_{edge}\leq 9$. For $N_{edge}=11$ the depolarization effect moves to a different filling.} 
\label{fig:fig8}
\end{figure}

Figure \ref{fig:fig8} shows spin phase diagrams for triangles with odd number
of degenerate edge states $N_{edge}$ and increasing size. Clearly, the total
spin depends on the filling factor and size of the triangle. However, all
charge-neutral systems at $\nu=1$ are always maximally spin polarized and a
complete depolarization occurs for $N_{edge}\leq 9$ for structures with one
extra electron added (such depolarization also occurs for even $N_{edge}$, not
shown). Similar results for small size triangles were obtained in our previous
work.\cite{Guclu+09} However, at $N_{edge}=11$ we do not observe
depolarization for $N_{edge}+1$ electrons but for $N_{edge}+3$, where a
formation of Wigner-like molecule for a triangle with $N_{edge}=7$ was
observed. We will come back to this problem later. We now focus on the
properties close to the charge neutrality.

For the charge-neutral case, the ground state corresponds to only one
configuration $|GS\rangle=\prod_i a_{i,\downarrow}^\dagger |0\rangle$ with maximum total $S_z$
and  occupation of all degenerate shell levels $i$ by electrons with parallel
spin. Here $|0\rangle$ is the HF ground state of all valence electrons.  Let us
consider the stability of the spin polarized state to single spin flips.  We
construct spin-flip excitations $|kl\rangle=a_{k,\uparrow}^\dagger  a_{l,\downarrow} |GS\rangle$ from the spin-polarized degenerate shell. The spin-up electron interacts
with a spin-down "hole'' in a spin-polarized state and forms a collective
excitation, an exciton. An exciton spectrum is obtained by building an exciton
Hamiltonian in the space of electron-hole pair excitations and diagonalizing
it numerically, as was done, e.g., for quantum dots.\cite{Hawrylak+Wojs+Brum+96} If the energy of the spin flip excitation turns out
to be negative in comparison with the spin-polarized ground state, the exciton
is bound and the spin-polarized state is unstable. The binding energy of a
spin-flip exciton is a difference between the energy of the lowest state with
$S=S_z^{max}-1$ and the energy of the spin-polarized ground state with
$S=S_z^{max}$. An advantage of this approach is  the ability to test the
stability of the spin polarized ground state for much larger TGQD sizes.

\begin{figure}
\epsfig{file=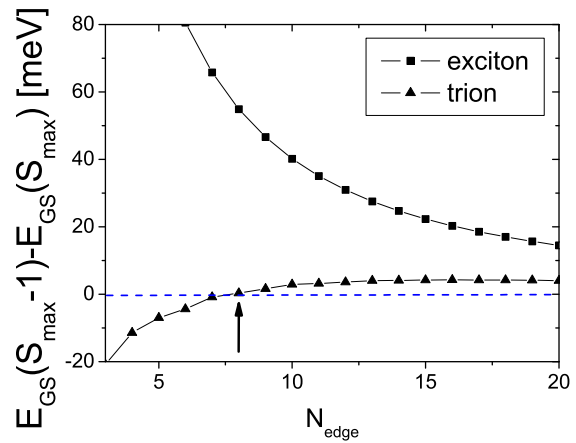,width=3.4in}
\caption{Size-dependent analysis based on exciton and trion binding
  energies. For the charge-neutral system, it is energetically unfavorable to
  form an exciton, which is characterized by a positive binding
  energy. Observed dependence confirms Lieb's theorem regarding the
  magnetization of the bipartite lattice systems. The formation of a trion is
  desirable for small size systems. The phase transition occurs close to
  $N_{edge}=8$, indicated by an arrow. The complete depolarization effects
  close to the charge neutrality observed previously in TGQD with
  $N_{edge}\leq 9$ for $N_{edge}+1$ electrons in Fig. \ref{fig:fig8} is
  predicted to not appear for larger systems.} 
\label{fig:fig9}
\end{figure}   
Figure \ref{fig:fig9} shows the exciton binding energy as a function of the
size of TGQD, labeled by a number of the degenerate states $N_{edge}$.  The
largest system, with $N_{edge}=20$, corresponds to a structure consisting of
$N=526$ atoms. The exciton binding energies are always positive, i.e., the
exciton does not form a bound state, confirming a stable magnetization of the
charge neutral system. The observed ferromagnetic order was also found by
other groups based on calculations for small systems with different levels of
approximations.\cite{Ezawa+07,FRP+07,Wang+Meng+08,Guclu+09} The above results
confirm predictions based on Lieb's theorem for a Hubbard model on bipartite
lattice relating total spin to the broken sublattice symmetry.\cite{Lieb+89} Unlike in Lieb's theorem, in our calculations many-body
interacting Hamiltonian contains direct long-range, exchange, and scattering
terms. Moreover, we include next-nearest-neighbor hopping integral in HF
self-consistent calculations that slightly violates bipartite lattice
property of the system, one of cornerstones of Lieb's arguments.\cite{Lieb+89}  Nevertheless, the main result of the spin-polarized ground
state for the charge neutral TGQD seems to be consistent with predictions of
Lieb's theorem and, hence, applicable to much larger systems. 

Having established the spin polarization of the charge-neutral TGQD we now
discuss the spin of charged TGQD. We start with a spin-polarized ground state $|GS\rangle$ of a charge-neutral TGQD with all electron spins down and add to it a
minority spin electron in any of the degenerate shell states $i$ as
$|i\rangle=a_{i,\uparrow}^\dagger|GS\rangle$. The total spin of these states is
$S_z^{max}-1/2$. We next study stability of such states with one minority spin-up electron to spin-flip excitations by forming three particle states
$|lki\rangle=a_{l,\uparrow}^\dagger  a_{k,\downarrow}a_{i,\uparrow}^\dagger |GS\rangle$ with total
spin $S_z^{max}-1/2-1$. Here there are two spin-up electrons and one hole with
spin-down in the spin-polarized ground state. The interaction between the two
electrons and a hole leads to the formation of trion states. We form a
Hamiltonian matrix in the space of three particle configurations and
diagonalize it to obtain trion states. If the energy of the lowest trion state
with $S_z^{max}-1/2-1 $ is lower  than the energy of any of the charged TGQD
states $|i\rangle$ with  $S_z^{max}-1/2$, the minority spin electron forms a bound state with the spin-flip exciton, a trion, and the spin-polarized state of a charged TGQD is unstable. The trion binding energy, shown in
Fig. \ref{fig:fig9}, is found to be negative for small systems with
$N_{edge}\leq 8$ and positive for all larger systems studied here. The binding
of the trion, i.e., the negative binding energy, is consistent with the complete
spin depolarization obtained using TB+HF+CI method for TGQD with $N_{edge}\leq
9$ but not observed for $N_{edge}=11$ (and not observed for $N_{edge}=10$, not
shown here), as shown in Fig. \ref{fig:fig8}. For small systems, a minority
spin-up electron triggers spin-flip excitations, which leads to the spin
depolarization. With increasing size, the effect of the correlations close to
the charge neutrality vanishes. At a critical size, around $N_{edge}=8$,
indicated by an arrow in Fig. \ref{fig:fig9}, a quantum phase transition
occurs from minimum to maximum total spin.

However, the spin depolarization does not vanish but moves to different
filling factors. In Fig. \ref{fig:fig8} we observe that the minimum spin state
for the largest structure computed by the TB+HF+CI method with $N_{edge}=11$
occurs for TGQD charged with additional three electrons. We recall that for
TGQD with $N_{edge}=7$  charged with three additional electrons a formation of
a Wigner-like spin polarized molecule was observed, shown in
Fig. \ref{fig:fig6}. In the following, the differences in the behavior of
these two systems, $N_{edge}=7$ and $11$, will be explained based on the
analysis of the many-body spectrum of the $N_{edge}=11$ system. 

\begin{figure}
\epsfig{file=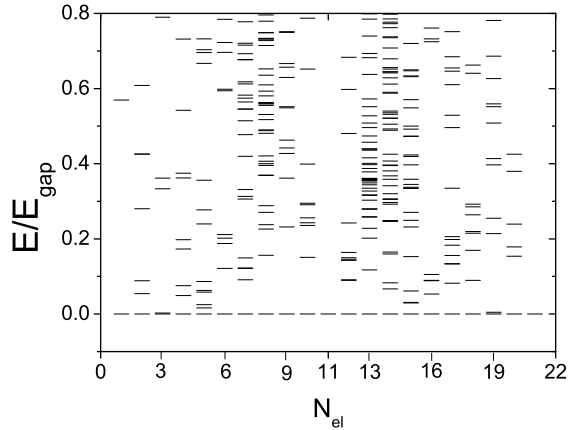,width=3.4in}
\caption{The low-energy spectra of the many-body states as a function of the
  number of electrons occupying the degenerate shell for the triangle with
  $N_{edge}=11$ degenerate states. The energies are renormalized by the energy
  gap corresponding to the half-filled shell, $N_{el}=11$ electrons. The large
  density of states related to the correlation effects observed in
  Fig. \ref{fig:fig4} around $N_{edge}+1$ electrons shifts to a different
  filling around $N_{edge}+3$ electrons.} 
\label{fig:fig10}
\end{figure}

Figure \ref{fig:fig10} shows the many-body energy spectra for different numbers
of electrons for $N_{edge}=11$ TGQD to be compared with Fig. \ref{fig:fig4}
for the $N_{edge}=7$ structure. Energies are renormalized by the energy gap of a
half-filled shell, $N_{el}=11$ electrons in this case. In contrast to the
$N_{edge}=7$ structure, energy levels corresponding to $N_{el}=N_{edge}+1$
electrons are sparse, whereas increased low-energy
densities of states appear for $N_{el}=N_{edge}+2$ and $N_{el}=N_{edge}+3$
electrons. In this structure, electrons are not as strongly confined as for
smaller systems. Therefore, for $N_{el}=N_{edge}+3$ electrons, geometrical
effects that lead to the formation of a Wigner-like molecule become less
important. Here, correlations dominate, which results in a large
low-energy density of states.

\section{IV. Different levels of approximations analysis}
In this section, we study the role of different interaction terms included in
our calculations. The computational procedure is identical to that described
in Sec. II. We start from the TB model but in  self-consistent HF and CI
calculations we include only specific Coulomb matrix elements. We compare
results obtained with Hubbard model with only the on-site term, the extended Hubbard model with on-site plus long range Coulomb interactions, and a model with all direct and exchange terms calculated for up to next-nearest neighbors using
Slater orbitals, and all longer range direct Coulomb interaction terms
approximated as $\langle ij|V|ji\rangle=1/(\kappa|r_i-r_j|)$, written in atomic units, 1 a.u.$=27.211$ eV, where $r_i$ and $r_j$ are positions of $i$-th and $j$-th sites,
respectively.

\begin{figure}
\epsfig{file=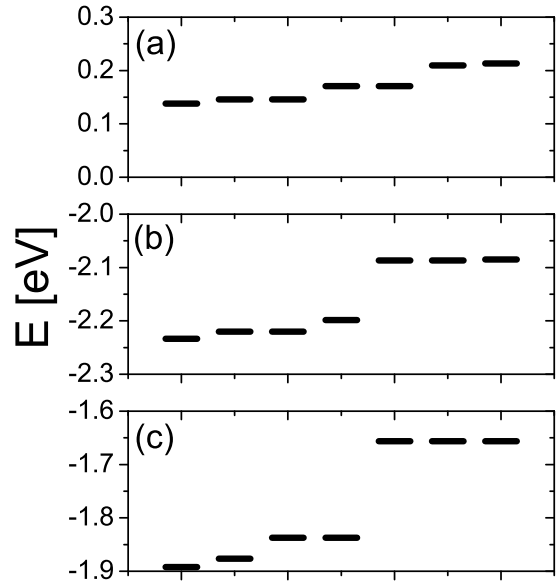,width=3.4in}
\caption{Hartree-Fock energy levels corresponding to the degenerate shell for
  calculations with (a) only the on-site term $U$ (Hubbard model), (b) the on-site term $U$ + direct long range interaction (extended Hubbard model), and (c) all interactions. A separation of three corner states with higher energies is related to direct long range Coulomb interaction terms.} 
\label{fig:fig11}
\end{figure}

The comparison of HF energy levels for the structure with $N_{edge}=7$ is
shown in Fig. \ref{fig:fig11}. 
The on-site $U$-term slightly removes degeneracy of the perfectly flat shell
[Fig.\ref{fig:fig11}(a)] and unveils the double valley degeneracy. On the
other hand, the direct long-range Coulomb interaction separates three corner
states from the rest with a higher energy [Fig.\ref{fig:fig11}(b)], forcing
the lifting of one of the doubly degenerate subshells. Finally, the inclusion
of exchange and scattering terms causes stronger removal of the degeneracy and
changes the order of the four lower-lying states. However, the form of the HF
orbitals is not affected significantly (not shown here).   

\begin{figure}
\epsfig{file=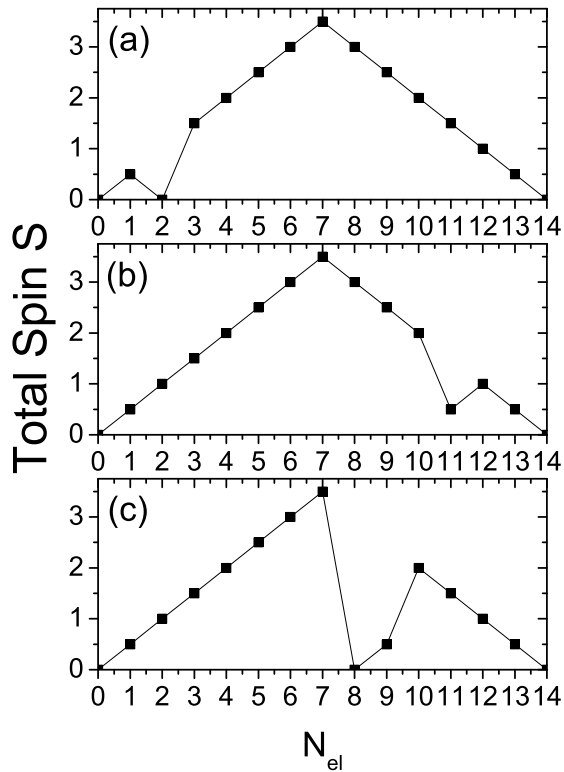,width=3.4in}
\caption{Spin phase diagrams obtained by use of the CI method with (a) only the on-site term $U$ (Hubbard model), (b) the on-site term $U$ + direct long range interaction (extended Hubbard model), and (c) all interactions. The ferromagnetic order for the charge-neutral system is properly predicted by all three methods. Correlations leading to the complete depolarization for $N_{el}=N_{edge}+1$ electrons and $N_{el}=N_{edge}+2$ electrons are observed only within a full interacting Hamiltonian.} 
\label{fig:fig12}
\end{figure}

In Fig. \ref{fig:fig12} we study the influence of different interacting terms
on CI results. The phase diagrams obtained within (a) the Hubbard model and (b) the extended Hubbard model show that all electronic phases are almost always fully spin polarized. The ferromagnetic order for the charge-neutral system is
properly predicted. For TGQD charged with electrons, only inclusion of all
Coulomb matrix elements correctly predicts the effect of the correlations
leading to the complete depolarization for $N_{el}=8$ and $9$. We note that
the depolarizations at other filling factors are also observed in Hubbard (at
$N_{el}=2)$) and extended Hubbard calculation (at $N_{el}=11)$) results.

\begin{figure}
\epsfig{file=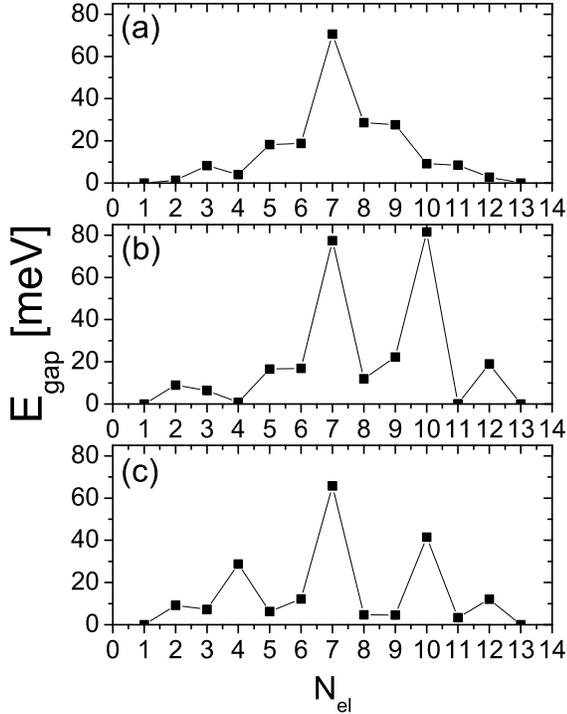,width=3.4in}
\caption{The excitation gaps corresponding to phase diagrams from
  Fig. \ref{fig:fig12} for many-body Hamiltonians with (a) only the on-site term $U$ (Hubbard model), (b) the on-site term $U$ + direct long range interaction (extended Hubbard model), and (c) all interactions. All three methods give qualitatively similar excitation gaps for the charge neutral system. A large energy gap for $N_{el}=N_{edge}+3$ electrons, which is related to geometrical properties of the structure, can be obtained by inclusion of direct long range interactions. This gap is slightly reduced by inclusion of exchange and scattering terms.} 
\label{fig:fig13}
\end{figure}

A more detailed analysis can be done by looking at the energy excitation gaps,
which are shown in Fig. \ref{fig:fig13}. For the charge-neutral system, all
three methods give comparable excitation gaps, in agreement with
previous results.\cite{Ezawa+07,FRP+07,Wang+Meng+08,Guclu+09} In the Hubbard
model, the energy gap of the doped system is reduced compared to the charge
neutrality but without affecting magnetic properties. The inclusion of a
direct long-range interaction in Fig. \ref{fig:fig13}(b) induces oscillations
of the energy gap. For $N_{el}=N_{edge}+1$ electrons the energy gap is
significantly reduced but the effect is not sufficiently strong to depolarize
the system. Further away from half-filling, a large energy gap for models with
long-range interactions for $N_{el}=N_{edge}+3$ appears, corresponding to the
formation of a Wigner-like molecule of three spin-polarized electrons in three
different corners. The inclusion of exchange and scattering terms slightly
reduces the gap but without changing a main effect of Wigner-like molecule
formation. 

\section{V. Conclusions and remarks}
We have investigated magnetism, correlations, and geometrical effects in TGQDs
by use of the TB+HF+CI method. Our many-body Hamiltonian includes all direct
long-range terms and exchange and scattering terms up to next-nearest
neighbors. We have performed analysis as a function of the filling factor of
the degenerate band of edge states for different sizes. Through a full
analysis of the many-body energy spectrum of structures consisting of up to
200 atoms, we confirmed the existence of the spin polarized ground state in
agreement  with  Lieb's theorem. By studying spin exciton binding energies, we
also predicted stable magnetization for structures with more than 500
atoms. The complete spin depolarization was observed for one  electron added
to the charge  neutral TGQD up to a critical size. Above a critical size the
maximally spin-polarized charged TGQD was predicted using trion binding energy
analysis. We have shown that in small systems, three electrons/holes added to
the charge neutrality form the spin-polarized Wigner-like molecule. We relate
this fact to geometrical effects and direct long-range interaction terms. For
larger systems, geometry becomes less important and for the same
filling we observe a spin depolarization as a result of 
correlations. Finally, we compared the fully interacting model with the Hubbard and extended Hubbard models. While qualitative agreement for the
charge-neutral system was observed, the effect of correlations can be described
only with the inclusion of all direct long-range, exchange, and scattering
terms. 
  
{\it Acknowledgment}. The authors thank NSERC, NRC-CNRS CRP, the Canadian Institute for Advanced Research, Institute for Microstructural Sciences, and QuantumWorks for support. P. P. acknowledges financial support from fellowship cofinanced by European Union within European Social Fund. A. W. acknowledges support from the EU Marie Curie CIG. 
 
\section{Appendix A}
In this appendix, we calculate density matrix elements $\rho_{jl\sigma'}^{o}$
between sites $j$ and $l$ for an infinite graphene sheet. The valence band
eigenfunctions of the TB Hamiltonian in the nearest-neighbor approximation given
by Eq. (\ref{HMF}) are
\\
\begin{eqnarray}
\nonumber
\Psi_{\bf k}^v\left({\bf r}\right)&=&\frac{1}{\sqrt{2N_c}}(
\sum_{\bf R_A}e^{i{\bf k}{\bf R_A}}\phi_{z}
\left({\bf r}-{\bf R_A}\right)
\\
&+&\sum_{{\bf{R_B}}}e^{i{\bf{k}}{\bf{R_B}}}e^{-i{\bf{k}}{\bf{b}}}
e^{-i\theta_{{\bf{k}}}}\phi_{z}\left({{\bf{r}}-{\bf{R_B}}}\right)),
\label{TBvalen}
\end{eqnarray}
where $\phi_{z}\left({{\bf{r}}}\right)$ are $p_{z}$ orbitals. The positions of
the sublattice A and B atoms are given by ${\bf R_A}=n{\bf{a}}_{1}+m{\bf{a}}_{2}$
and ${\bf R_B}=n{\bf{a}}_{1}+m{\bf{a}}_{2}+{\bf b}$, described by unit vectors of
hexagonal lattice defined as ${\bf{a}}_{1,2}=a/2(\pm \sqrt{3},3)$ and
${\bf{b}}=a(0,1)$, a vector between two nearest-neighbors atoms from the same
unit cell with a distance $a=1.42$  $\rm{\AA}$. $N_{c}$ is the number of unit
cells, and $\exp({i\theta_{\bf k}})=\frac{f({\bf{k}})}{|f({\bf{k}})|}$ with
$f({\bf{k}})=1+e^{i{\bf{k}}{\bf{a}}_{1}}+e^{i{\bf{k}}{\bf{a}}_{2}}$. The density
matrix for the graphene layer $\rho^{o}_{jl\sigma}$ for two sites $j$ and $l$ is defined as 
\begin{eqnarray}
\rho^{o}_{jl\sigma}=\sum_{\bf k}
b_{{\bf R}_j}({\bf k})b_{{\bf R}_l}({\bf k}),
\end{eqnarray}  
where $b_{\bf R}$'s are the coefficients of the $p_z$ orbitals given in Eq. (\ref{TBvalen}). The on-site density matrix element for an arbitrary lattice site {\it j} is site and sublattice index independent,
\begin{eqnarray}
\rho^{o}_{jj\sigma}=\frac{1}{2N_{c}}\sum_{{\bf{k}}}
e^{-i{\bf k R}_j}e^{i{\bf k R}_j}=
\frac{1}{2N_{c}}\sum_{{\bf{k}}}1=\frac{1}{2},
\label{fk}
\end{eqnarray} 
where we took into account the fact that the number of occupied states is
equal to the number of unit cells in the system. The nearest-neighbors density
matrix elements for two atoms from the same unit cell corresponds to ${\bf R}_l-{\bf R}_j={\bf b}$ and can be calculated using 
\begin{eqnarray}
\nonumber
\rho^{o}_{jl\sigma}&=&
\frac{1}{2N_{c}}\sum_{{\bf{k}}}e^{-i{\bf k R}_j}
e^{i {\bf k R}_l} e^{-i{\bf k b}}e^{-i\theta_{\bf k}}
\\
&=&\frac{1}{2N_{c}}\sum_{{\bf{k}}}e^{-i\theta_{{\bf{k}}}}\simeq 0.262, 
\nonumber
\end{eqnarray}
where the summation over occupied valence states is carried out numerically. We obtained the same value for two other nearest neighbors. The same results can also be obtained by diagonalizing a sufficiently large hexagonal
graphene quantum dot and  by computing the density matrix elements for two
nearest neighbors in the vicinity of the center of the structure. We have also
calculated next-nearest neighbors density matrix elements, obtaining negligibly
small values.



\begin{thebibliography}{29}
\expandafter\ifx\csname natexlab\endcsname\relax\def\natexlab#1{#1}\fi
\expandafter\ifx\csname bibnamefont\endcsname\relax
  \def\bibnamefont#1{#1}\fi
\expandafter\ifx\csname bibfnamefont\endcsname\relax
  \def\bibfnamefont#1{#1}\fi
\expandafter\ifx\csname citenamefont\endcsname\relax
  \def\citenamefont#1{#1}\fi
\expandafter\ifx\csname url\endcsname\relax
  \def\url#1{\texttt{#1}}\fi
\expandafter\ifx\csname urlprefix\endcsname\relax\def\urlprefix{URL }\fi
\providecommand{\bibinfo}[2]{#2}
\providecommand{\eprint}[2][]{\url{#2}}

\bibitem[{\citenamefont{Novoselov et~al.}(2004)\citenamefont{Novoselov, Geim,
  Morozov, Jiang, Zhang, Dubonos, Grigorieva, and Firsov}}]{Novoselov+Geim+04}
\bibinfo{author}{\bibfnamefont{K.~S.} \bibnamefont{Novoselov}},
  \bibinfo{author}{\bibfnamefont{A.~K.} \bibnamefont{Geim}},
  \bibinfo{author}{\bibfnamefont{S.~V.} \bibnamefont{Morozov}},
  \bibinfo{author}{\bibfnamefont{D.}~\bibnamefont{Jiang}},
  \bibinfo{author}{\bibfnamefont{Y.}~\bibnamefont{Zhang}},
  \bibinfo{author}{\bibfnamefont{S.~V.} \bibnamefont{Dubonos}},
  \bibinfo{author}{\bibfnamefont{I.~V.} \bibnamefont{Grigorieva}},
  \bibnamefont{and} \bibinfo{author}{\bibfnamefont{A.~A.}
  \bibnamefont{Firsov}}, \bibinfo{journal}{Science}
  \textbf{\bibinfo{volume}{306}}, \bibinfo{pages}{666} (\bibinfo{year}{2004}).

\bibitem[{\citenamefont{Novoselov et~al.}(2005)\citenamefont{Novoselov, Geim,
  Morozov, Jiang, Katsnelson, Grigorieva, Dubonos, and Firsov}}]{Novoselov+Geim+05}
\bibinfo{author}{\bibfnamefont{K.~S.} \bibnamefont{Novoselov}},
  \bibinfo{author}{\bibfnamefont{A.~K.} \bibnamefont{Geim}},
  \bibinfo{author}{\bibfnamefont{S.~V.} \bibnamefont{Morozov}},
  \bibinfo{author}{\bibfnamefont{D.}~\bibnamefont{Jiang}},
  \bibinfo{author}{\bibfnamefont{M.~I.} \bibnamefont{Katsnelson}},
  \bibinfo{author}{\bibfnamefont{I.~V.} \bibnamefont{Grigorieva}},
  \bibinfo{author}{\bibfnamefont{S.~V.} \bibnamefont{Dubonos}},
  \bibnamefont{and} \bibinfo{author}{\bibfnamefont{A.~A.}
  \bibnamefont{Firsov}}, \bibinfo{journal}{Nature}
  \textbf{\bibinfo{volume}{438}}, \bibinfo{pages}{197} (\bibinfo{year}{2005}).

\bibitem[{\citenamefont{Zhang et~al.}(2005)\citenamefont{Zhang, Tan, Stormer,
  and Kim}}]{Zhang+Tan+05}
\bibinfo{author}{\bibfnamefont{Y.~B.} \bibnamefont{Zhang}},
  \bibinfo{author}{\bibfnamefont{Y.~W.} \bibnamefont{Tan}},
  \bibinfo{author}{\bibfnamefont{H.~L.} \bibnamefont{Stormer}},
  \bibnamefont{and} \bibinfo{author}{\bibfnamefont{P.}~\bibnamefont{Kim}},
  \bibinfo{journal}{Nature} \textbf{\bibinfo{volume}{438}},
  \bibinfo{pages}{201} (\bibinfo{year}{2005}).

\bibitem[{\citenamefont{Son et~al.}(2006)\citenamefont{Son, Cohen, and Louie}}]{Son+PRL+06}
\bibinfo{author}{\bibfnamefont{Y.~W.} \bibnamefont{Son}},
  \bibinfo{author}{\bibfnamefont{M.~L.} \bibnamefont{Cohen}}, \bibnamefont{and}
  \bibinfo{author}{\bibfnamefont{S.~G.} \bibnamefont{Louie}},
  \bibinfo{journal}{Phys. Rev. Lett.} \textbf{\bibinfo{volume}{97}},
  \bibinfo{pages}{216803} (\bibinfo{year}{2006}).

\bibitem[{\citenamefont{Sadowski et~al.}(2006)\citenamefont{Sadowski, Martinez, Potemski, Berger, and de Heer}}]{Potemski+deHeer+06}
\bibinfo{author}{\bibfnamefont{M.~L.} \bibnamefont{Sadowski}},
  \bibinfo{author}{\bibfnamefont{G.} \bibnamefont{Martinez}},
  \bibinfo{author}{\bibfnamefont{M.} \bibnamefont{Potemski}},
  \bibinfo{author}{\bibfnamefont{C.} \bibnamefont{Berger}},
\bibnamefont{and}
  \bibinfo{author}{\bibfnamefont{W.~A.} \bibnamefont{de Heer}},
  \bibinfo{journal}{Phys. Rev. Lett.} \textbf{\bibinfo{volume}{97}},
  \bibinfo{pages}{266405} (\bibinfo{year}{2006}).

\bibitem[{\citenamefont{Geim et~al.}(2007)\citenamefont{Geim, Novoselov}}]{Geim+Novoselov+07}
\bibinfo{author}{\bibfnamefont{A.~K.} \bibnamefont{Geim}}
\bibnamefont{and}
  \bibinfo{author}{\bibfnamefont{K.~S.} \bibnamefont{Novoselov}},
  \bibinfo{journal}{Nat. Mater.} \textbf{\bibinfo{volume}{6}},
  \bibinfo{pages}{183} (\bibinfo{year}{2007}). 

\bibitem[{\citenamefont{Rycerz et~al.}(2007)\citenamefont{Rycerz, Tworzydlo, and Beenakker}}]{Rycerz+Tworzydlo+07}
\bibinfo{author}{\bibfnamefont{A.} \bibnamefont{Rycerz}},
  \bibinfo{author}{\bibfnamefont{J.}~\bibnamefont{Tworzydlo}}, \bibnamefont{and}
  \bibinfo{author}{\bibfnamefont{C.~W.} \bibnamefont{Beenakker}},
  \bibinfo{journal}{Nature Phys.} \textbf{\bibinfo{volume}{3}},
  \bibinfo{pages}{172} (\bibinfo{year}{2007}). 

\bibitem[{\citenamefont{Xia et~al.}(2009)\citenamefont{Xia, Mueller, Lin, Valdes-Garcia and Avouris}}]{Xia+Mueller+09}
\bibinfo{author}{\bibfnamefont{F.}~\bibnamefont{Xia}}, 
  \bibinfo{author}{\bibfnamefont{T.} \bibnamefont{Mueller}},
  \bibinfo{author}{\bibfnamefont{Y.-M.} \bibnamefont{Lin}},
  \bibinfo{author}{\bibfnamefont{A.} \bibnamefont{Valdes-Garcia}},
  \bibnamefont{and}
  \bibinfo{author}{\bibfnamefont{P.} \bibnamefont{Avouris}},
  \bibinfo{journal}{Nat. Nanotechnol.} \textbf{\bibinfo{volume}{4}},
  \bibinfo{pages}{839} (\bibinfo{year}{2009}). 

\bibitem[{\citenamefont{Mueller et~al.}(2010)\citenamefont{Mueller, Xia, and Avouris}}]{Mueller+Xia+10}
\bibinfo{author}{\bibfnamefont{T.} \bibnamefont{Mueller}},
  \bibinfo{author}{\bibfnamefont{F.}~\bibnamefont{Xia}}, \bibnamefont{and}
  \bibinfo{author}{\bibfnamefont{P.} \bibnamefont{Avouris}},
  \bibinfo{journal}{Nat. Photon.} \textbf{\bibinfo{volume}{4}},
  \bibinfo{pages}{297} (\bibinfo{year}{2010}).

\bibitem[{\citenamefont{Neto et~al.}(2009)\citenamefont{Neto, Guinea, Peres,
  Novoselov, and Geim}}]{Neto+Guinea+09}
\bibinfo{author}{\bibfnamefont{A.~H.~C.} \bibnamefont{Neto}},
  \bibinfo{author}{\bibfnamefont{F.}~\bibnamefont{Guinea}},
  \bibinfo{author}{\bibfnamefont{N.~M.~R.} \bibnamefont{Peres}},
  \bibinfo{author}{\bibfnamefont{K.~S.} \bibnamefont{Novoselov}},
  \bibnamefont{and} \bibinfo{author}{\bibfnamefont{A.~K.} \bibnamefont{Geim}},
  \bibinfo{journal}{Rev. of Mod. Phys.} \textbf{\bibinfo{volume}{81}},
  \bibinfo{pages}{109} (\bibinfo{year}{2009}). 

\bibitem[{\citenamefont{Rozhkov et~al.}(2011)\citenamefont{Rozhkov, Giavaras, Bliokh, Freilikher, and Nori}}]{Rozhkov+11}
\bibinfo{author}{\bibfnamefont{A.~H.} \bibnamefont{Rozhkov}}, 
  \bibinfo{author}{\bibfnamefont{G.} \bibnamefont{Giavaras}}, 
  \bibinfo{author}{\bibfnamefont{Y.~P.} \bibnamefont{Bliokh}},
  \bibinfo{author}{\bibfnamefont{V.} \bibnamefont{Freilikher}},
 \bibnamefont{and}
  \bibinfo{author}{\bibfnamefont{F.}~\bibnamefont{Nori}},
  \bibinfo{journal}{Phys. Rep.} \textbf{\bibinfo{volume}{81}}
  , \bibinfo{pages}{195414} (\bibinfo{year}{2011}).

\bibitem[{\citenamefont{Abergel et~al.}(2010)\citenamefont{Abergel, Apalkov, Berashevich, Ziegler, and Chakraborty}}]{Abergel+10}
\bibinfo{author}{\bibfnamefont{A.~H.} \bibnamefont{Abergel}}, 
  \bibinfo{author}{\bibfnamefont{G.} \bibnamefont{Apalkov}}, 
  \bibinfo{author}{\bibfnamefont{Y.~P.} \bibnamefont{Berashevich}},
  \bibinfo{author}{\bibfnamefont{V.} \bibnamefont{Ziegler}},
 \bibnamefont{and}
  \bibinfo{author}{\bibfnamefont{F.}~\bibnamefont{Chakraborty}},
  \bibinfo{journal}{Adv. in Phys.} \textbf{\bibinfo{volume}{59}}
  , \bibinfo{pages}{261} (\bibinfo{year}{2010}).

\bibitem[{\citenamefont{Li et~al.}(2008)\citenamefont{Li, Wang, Zhang, Lee, and Dai}}]{Li+08}
\bibinfo{author}{\bibfnamefont{X.} \bibnamefont{Li}},
  \bibinfo{author}{\bibfnamefont{X.} \bibnamefont{Wang}},
  \bibinfo{author}{\bibfnamefont{L.} \bibnamefont{Zhang}},
  \bibinfo{author}{\bibfnamefont{S.} \bibnamefont{Lee}}, \bibnamefont{and}
  \bibinfo{author}{\bibfnamefont{H.} \bibnamefont{Dai}},
  \bibinfo{journal}{Science} \textbf{\bibinfo{volume}{319}},
  \bibinfo{pages}{1229} (\bibinfo{year}{2008}).

\bibitem[{\citenamefont{Ponomarenko et~al.}(2008)\citenamefont{Ponomarenko, Schedin, Katsnelson, Yang, Hill, Novoselov, and Geim}}]{Ponomarenko+08}
\bibinfo{author}{\bibfnamefont{L.~A.} \bibnamefont{Ponomarenko}},
  \bibinfo{author}{\bibfnamefont{F.} \bibnamefont{Schedin}},
  \bibinfo{author}{\bibfnamefont{M.~I.} \bibnamefont{Katsnelson}},
  \bibinfo{author}{\bibfnamefont{R.} \bibnamefont{Yang}},
  \bibinfo{author}{\bibfnamefont{E.~W.} \bibnamefont{Hill}},
  \bibinfo{author}{\bibfnamefont{K.~S.} \bibnamefont{Novoselov}}, \bibnamefont{and}
  \bibinfo{author}{\bibfnamefont{A.~K.} \bibnamefont{Geim}},
  \bibinfo{journal}{Science} \textbf{\bibinfo{volume}{320}},
  \bibinfo{pages}{356} (\bibinfo{year}{2008}).

\bibitem[{\citenamefont{Ci et~al.}(2008)\citenamefont{Ci, Xu, Wang, Gao, Ding, Kelly, and Dai}}]{Ci+08}
\bibinfo{author}{\bibfnamefont{L.} \bibnamefont{Ci}},
  \bibinfo{author}{\bibfnamefont{Z.} \bibnamefont{Xu}},
  \bibinfo{author}{\bibfnamefont{L.} \bibnamefont{Wang}},
  \bibinfo{author}{\bibfnamefont{W.} \bibnamefont{Gao}},
  \bibinfo{author}{\bibfnamefont{F.} \bibnamefont{Ding}},
  \bibinfo{author}{\bibfnamefont{K.~F.} \bibnamefont{Kelly}},
  \bibinfo{author}{\bibfnamefont{B.~I.} \bibnamefont{Yakobson}}, \bibnamefont{and}
  \bibinfo{author}{\bibfnamefont{P.~M.} \bibnamefont{Ajayan}},
  \bibinfo{journal}{Nano Res} \textbf{\bibinfo{volume}{1}},
  \bibinfo{pages}{116} (\bibinfo{year}{2008}).

\bibitem[{\citenamefont{You et~al.}(2008)\citenamefont{You, Ni, Yu, and Shena}}]{You+08}
\bibinfo{author}{\bibfnamefont{Y.} \bibnamefont{You}},
  \bibinfo{author}{\bibfnamefont{Z.} \bibnamefont{Ni}},
  \bibinfo{author}{\bibfnamefont{T.} \bibnamefont{Yu}},
 \bibnamefont{and}
  \bibinfo{author}{\bibfnamefont{Z.} \bibnamefont{Shena}},
  \bibinfo{journal}{Appl. Phys. Lett.} \textbf{\bibinfo{volume}{93}},
  \bibinfo{pages}{163112} (\bibinfo{year}{2008}).

\bibitem[{\citenamefont{Schnez et~al.}(2009)\citenamefont{Schnez, Molitor, Stampfer, G\"{u}ttinger, Shorubalko, Ihn, and Ensslin}}]{Schnez+09}
\bibinfo{author}{\bibfnamefont{S.} \bibnamefont{Schnez}},
  \bibinfo{author}{\bibfnamefont{F.} \bibnamefont{Molitor}},
  \bibinfo{author}{\bibfnamefont{C.} \bibnamefont{Stampfer}},
  \bibinfo{author}{\bibfnamefont{J.} \bibnamefont{G\"{u}ttinger}},
  \bibinfo{author}{\bibfnamefont{I.} \bibnamefont{Shorubalko}},
  \bibinfo{author}{\bibfnamefont{T.} \bibnamefont{Ihn}},
 \bibnamefont{and}
  \bibinfo{author}{\bibfnamefont{K.} \bibnamefont{Ensslin}},
  \bibinfo{journal}{Appl. Phys. Lett.} \textbf{\bibinfo{volume}{94}},
  \bibinfo{pages}{012107} (\bibinfo{year}{2009}).

\bibitem[{\citenamefont{Ritter and Lyding}(2009)}]{Ritter+09} 
\bibinfo{author}{\bibfnamefont{K.~A.}~\bibnamefont{Ritter}}
  \bibnamefont{and} \bibinfo{author}{\bibfnamefont{J.~W.}
  \bibnamefont{Lyding}}, \bibinfo{journal}{Nat Mater.}
  \textbf{\bibinfo{volume}{8}}, \bibinfo{pages}{235} (\bibinfo{year}{2009}).


\bibitem[{\citenamefont{Jia et~al.}(2009)\citenamefont{Jia, Hofmann, Meunier, Sumpter, Campos-Delgado, Romo-Herrera, Son, Hsieh, Reina, Kong, Terrones, and Dresselhaus}}]{Jia+09}
\bibinfo{author}{\bibfnamefont{X.} \bibnamefont{Jia}},
  \bibinfo{author}{\bibfnamefont{M.} \bibnamefont{Hofmann}},
  \bibinfo{author}{\bibfnamefont{V.} \bibnamefont{Meunier}},
  \bibinfo{author}{\bibfnamefont{B.~G.} \bibnamefont{Sumpter}},
  \bibinfo{author}{\bibfnamefont{J.} \bibnamefont{Campos-Delgado}},
  \bibinfo{author}{\bibfnamefont{J.~M.} \bibnamefont{Romo-Herrera}},
  \bibinfo{author}{\bibfnamefont{H.} \bibnamefont{Son}},
  \bibinfo{author}{\bibfnamefont{Y.-P.} \bibnamefont{Hsieh}},
  \bibinfo{author}{\bibfnamefont{A.} \bibnamefont{Reina}},
  \bibinfo{author}{\bibfnamefont{J.} \bibnamefont{Kong}},
  \bibinfo{author}{\bibfnamefont{M.} \bibnamefont{Terrones}}, \bibnamefont{and}
  \bibinfo{author}{\bibfnamefont{M.~S.} \bibnamefont{Dresselhaus}},
  \bibinfo{journal}{Science} \textbf{\bibinfo{volume}{323}},
  \bibinfo{pages}{1701} (\bibinfo{year}{2009}).

\bibitem[{\citenamefont{Campos et~al.}(2009)\citenamefont{Campos, Manfrinato, Sanchez-Yamagishi, Kong, and Jarillo-Herrero}}]{Campos+09}
\bibinfo{author}{\bibfnamefont{L.~C.} \bibnamefont{Campos}},
  \bibinfo{author}{\bibfnamefont{V.~R.} \bibnamefont{Manfrinato}},
  \bibinfo{author}{\bibfnamefont{J.~D.} \bibnamefont{Sanchez-Yamagishi}},
  \bibinfo{author}{\bibfnamefont{J.} \bibnamefont{Kong}}, \bibnamefont{and}
  \bibinfo{author}{\bibfnamefont{P.} \bibnamefont{Jarillo-Herrero}},
  \bibinfo{journal}{Nano Lett.} \textbf{\bibinfo{volume}{9}},
  \bibinfo{pages}{2600} (\bibinfo{year}{2009}).

\bibitem[{\citenamefont{Neubeck et~al.}(2010)\citenamefont{Neubeck, You, Ni, Blake, Shen, Geim and Novoselov}}]{Neubeck+10}
\bibinfo{author}{\bibfnamefont{S.} \bibnamefont{Neubeck}},
  \bibinfo{author}{\bibfnamefont{Y.~M.} \bibnamefont{You}},
  \bibinfo{author}{\bibfnamefont{Z.~H.} \bibnamefont{Ni}},
  \bibinfo{author}{\bibfnamefont{P.} \bibnamefont{Blake}},
  \bibinfo{author}{\bibfnamefont{Z.~X.} \bibnamefont{Shen}},
  \bibinfo{author}{\bibfnamefont{A.~K.} \bibnamefont{Geim}}, \bibnamefont{and}
  \bibinfo{author}{\bibfnamefont{K.~S.} \bibnamefont{Novoselov}},
  \bibinfo{journal}{Appl. Phys. Lett.} \textbf{\bibinfo{volume}{97}},
  \bibinfo{pages}{053110} (\bibinfo{year}{2010}).

\bibitem[{\citenamefont{Bir$\acute{o}$ and Lambin}(2010)}]{Biro+10} 
\bibinfo{author}{\bibfnamefont{L.~P.}~\bibnamefont{Bir$\acute{o}}$}
  \bibnamefont{and} \bibinfo{author}{\bibfnamefont{Ph.}
  \bibnamefont{Lambin}}, \bibinfo{journal}{Carbon}
  \textbf{\bibinfo{volume}{48}}, \bibinfo{pages}{2677} (\bibinfo{year}{2010}).
 
\bibitem[{\citenamefont{Cruz-Silva et~al.}(2010)\citenamefont{Cruz-Silva, Botello-Mendez, Barnett, Blake, Jia, Dresselhaus, Terrones, Terrones, Sumpter and Meunier}}]{CruzSilva+10}
\bibinfo{author}{\bibfnamefont{E.} \bibnamefont{Cruz-Silva}},
  \bibinfo{author}{\bibfnamefont{A.~R.} \bibnamefont{Botello-Mendez}},
  \bibinfo{author}{\bibfnamefont{Z.~M.} \bibnamefont{Barnett}},
  \bibinfo{author}{\bibfnamefont{X.} \bibnamefont{Jia}},
  \bibinfo{author}{\bibfnamefont{M.~S.} \bibnamefont{Dresselhaus}},
  \bibinfo{author}{\bibfnamefont{H.} \bibnamefont{Terrones}},
  \bibinfo{author}{\bibfnamefont{M.} \bibnamefont{Terrones}},
  \bibinfo{author}{\bibfnamefont{B.~G.} \bibnamefont{Sumpter}}, \bibnamefont{and}
  \bibinfo{author}{\bibfnamefont{V.} \bibnamefont{Meunier}},
  \bibinfo{journal}{Phys. Rev. Lett.} \textbf{\bibinfo{volume}{105}},
  \bibinfo{pages}{045501} (\bibinfo{year}{2010}).

\bibitem[{\citenamefont{Yang et~al.}(2010)\citenamefont{Yang, Zhang, Wang, Shi, Shi, and Dai}}]{Yang+10}
\bibinfo{author}{\bibfnamefont{R.} \bibnamefont{Yang}},
  \bibinfo{author}{\bibfnamefont{L.} \bibnamefont{Zhang}},
  \bibinfo{author}{\bibfnamefont{Y.} \bibnamefont{Wang}},
  \bibinfo{author}{\bibfnamefont{Z.} \bibnamefont{Shi}},
  \bibinfo{author}{\bibfnamefont{D.} \bibnamefont{Shi}},
  \bibinfo{author}{\bibfnamefont{H.} \bibnamefont{Gao}},
  \bibinfo{author}{\bibfnamefont{E.} \bibnamefont{Wang}}, \bibnamefont{and}
  \bibinfo{author}{\bibfnamefont{G.} \bibnamefont{Zhang}},
  \bibinfo{journal}{Adv. Mater.} \textbf{\bibinfo{volume}{22}},
  \bibinfo{pages}{4014} (\bibinfo{year}{2010}).

\bibitem[{\citenamefont{Krauss et~al.}(2010)\citenamefont{Krauss, Nemes-Incze, Skakalova, Bir$\acute{o}$, von Klitzing, and Smet}}]{Krauss+10}
\bibinfo{author}{\bibfnamefont{B.} \bibnamefont{Krauss}},
  \bibinfo{author}{\bibfnamefont{P.} \bibnamefont{Nemes-Incze}},
  \bibinfo{author}{\bibfnamefont{V.} \bibnamefont{Skakalova}},
  \bibinfo{author}{\bibfnamefont{L.~P.} \bibnamefont{Bir$\acute{o}$}},
  \bibinfo{author}{\bibfnamefont{K.} \bibnamefont{von Klitzing}}, \bibnamefont{and}
  \bibinfo{author}{\bibfnamefont{J.~H.} \bibnamefont{Smet}},
  \bibinfo{journal}{Nano Lett.} \textbf{\bibinfo{volume}{10}},
  \bibinfo{pages}{4544} (\bibinfo{year}{2010}).

\bibitem[{\citenamefont{Zhi and M\"{u}llen}(2009)}]{Zhi+08} 
\bibinfo{author}{\bibfnamefont{L.}~\bibnamefont{Zhi}}
  \bibnamefont{and} \bibinfo{author}{\bibfnamefont{K.}
  \bibnamefont{M\"{u}llen}}, \bibinfo{journal}{J. Mater. Chem.}
  \textbf{\bibinfo{volume}{18}}, \bibinfo{pages}{1472} (\bibinfo{year}{2008}).

\bibitem[{\citenamefont{Treier et~al.}(2010)\citenamefont{Treier, Pignedoli, Laino, Rieger, M\"{u}llen, Passerone, and Fasel}}]{Treier+10} 
\bibinfo{author}{\bibfnamefont{M.}~\bibnamefont{Treier}},
  \bibinfo{author}{\bibfnamefont{C.~A.} \bibnamefont{Pignedoli}}, 
  \bibinfo{author}{\bibfnamefont{T.} \bibnamefont{Laino}},
\bibinfo{author}{\bibfnamefont{R.} \bibnamefont{Rieger}},
\bibinfo{author}{\bibfnamefont{K.} \bibnamefont{M\"{u}llen}},
\bibinfo{author}{\bibfnamefont{D.} \bibnamefont{Passerone}},
  \bibnamefont{and} \bibinfo{author}{\bibfnamefont{R.}
  \bibnamefont{Fasel}}, \bibinfo{journal}{Nat. Chem.}
  \textbf{\bibinfo{volume}{3}}, \bibinfo{pages}{61} (\bibinfo{year}{2010}).

\bibitem[{\citenamefont{Mueller et~al.}(2010)\citenamefont{Mueller, Yan, McGuire, and Li}}]{Mueller+10} 
\bibinfo{author}{\bibfnamefont{M.~L.}~\bibnamefont{Mueller}},
  \bibinfo{author}{\bibfnamefont{X.} \bibnamefont{Yan}}, 
  \bibinfo{author}{\bibfnamefont{J.~A.} \bibnamefont{McGuire}},
  \bibnamefont{and} \bibinfo{author}{\bibfnamefont{L.}
  \bibnamefont{Li}}, \bibinfo{journal}{Nano Lett.}
  \textbf{\bibinfo{volume}{10}}, \bibinfo{pages}{2679} (\bibinfo{year}{2010}).

\bibitem[{\citenamefont{Morita et~al.}(2011)\citenamefont{Morita, Suzuki, Sato, and Takui}}]{Morita+11} 
\bibinfo{author}{\bibfnamefont{Y.}~\bibnamefont{Morita}},
  \bibinfo{author}{\bibfnamefont{S.} \bibnamefont{Suzuki}}, 
  \bibinfo{author}{\bibfnamefont{K.} \bibnamefont{Sato}},
  \bibnamefont{and} \bibinfo{author}{\bibfnamefont{T.}
  \bibnamefont{Takui}}, \bibinfo{journal}{Nat. Chem.}
  \textbf{\bibinfo{volume}{3}}, \bibinfo{pages}{197} (\bibinfo{year}{2011}).

\bibitem[{\citenamefont{Lu et~al.}(2011)\citenamefont{Lu, Yeo, Gan, Wu, and Loh}}]{Lu+11} 
\bibinfo{author}{\bibfnamefont{J.}~\bibnamefont{Lu}},
  \bibinfo{author}{\bibfnamefont{P.~S.~E.} \bibnamefont{Yeo}}, 
  \bibinfo{author}{\bibfnamefont{C.~K.} \bibnamefont{Gan}},
  \bibinfo{author}{\bibfnamefont{P.} \bibnamefont{Wu}},
  \bibnamefont{and} \bibinfo{author}{\bibfnamefont{K.~P.}
  \bibnamefont{Loh}}, \bibinfo{journal}{Nat. Nanotechnol.}
  \textbf{\bibinfo{volume}{6}}, \bibinfo{pages}{247} (\bibinfo{year}{2011}).

\bibitem[{\citenamefont{Singh and Yakobson}(2009)}]{Singh+09} 
\bibinfo{author}{\bibfnamefont{A.~K.}~\bibnamefont{Singh}}
  \bibnamefont{and} \bibinfo{author}{\bibfnamefont{B.~I.}
  \bibnamefont{Yakobson}}, \bibinfo{journal}{Nano Lett.}
  \textbf{\bibinfo{volume}{9}}, \bibinfo{pages}{1540} (\bibinfo{year}{2009}).

\bibitem[{\citenamefont{Tozzini and Pellegrini}(2010)}]{Tozzini+10} 
\bibinfo{author}{\bibfnamefont{V.}~\bibnamefont{Tozzini}}
  \bibnamefont{and} \bibinfo{author}{\bibfnamefont{V.}
  \bibnamefont{Pellegrini}}, \bibinfo{journal}{Phys. Rev. B}
  \textbf{\bibinfo{volume}{81}}, \bibinfo{pages}{113404} (\bibinfo{year}{2010}).

\bibitem[{\citenamefont{Xiang et~al.}(2009)\citenamefont{Xiang, Kan, Wei, Whangbo, and Yang}}]{Xiang+09} 
\bibinfo{author}{\bibfnamefont{H.}~\bibnamefont{Xiang}},
  \bibinfo{author}{\bibfnamefont{E.} \bibnamefont{Kan}}, 
  \bibinfo{author}{\bibfnamefont{S.-H.} \bibnamefont{Wei}},
  \bibinfo{author}{\bibfnamefont{M.-H.} \bibnamefont{Whangbo}},
  \bibnamefont{and} \bibinfo{author}{\bibfnamefont{J.}
  \bibnamefont{Yang}}, \bibinfo{journal}{Nano Lett.}
  \textbf{\bibinfo{volume}{9}}, \bibinfo{pages}{4025} (\bibinfo{year}{2009}).

\bibitem[{\citenamefont{Schmidt and Loss}(2010)}]{Schmidt+10} 
\bibinfo{author}{\bibfnamefont{M.~J.}~\bibnamefont{Schmidt}}
  \bibnamefont{and} \bibinfo{author}{\bibfnamefont{D.}
  \bibnamefont{Loss}}, \bibinfo{journal}{Phys. Rev. B}
  \textbf{\bibinfo{volume}{82}}, \bibinfo{pages}{085422} (\bibinfo{year}{2010}).

\bibitem[{\citenamefont{Yamamoto et~al.}(2006)\citenamefont{Yamamoto,
  Noguchi, and Watanabe}}]{Yamamoto+06}
\bibinfo{author}{\bibfnamefont{T.}~\bibnamefont{Yamamoto}},
  \bibinfo{author}{\bibfnamefont{T.}~\bibnamefont{Noguchi}}, \bibnamefont{and}
  \bibinfo{author}{\bibfnamefont{K.}~\bibnamefont{Watanabe}},
  \bibinfo{journal}{Phys. Rev. B} \textbf{\bibinfo{volume}{74}},
  \bibinfo{pages}{121409} (\bibinfo{year}{2006}).

\bibitem[{\citenamefont{Zhang et~al.}(2008)\citenamefont{Zhang, Chang, and Peeters}}]{Zhang+08}
\bibinfo{author}{\bibfnamefont{Z.~Z.} \bibnamefont{Zhang}},
\bibinfo{author}{\bibfnamefont{K.} \bibnamefont{Chang}}, \bibnamefont{and}
  \bibinfo{author}{\bibfnamefont{F.~M.}~\bibnamefont{Peeters}},
  \bibinfo{journal}{Phys. Rev. B} \textbf{\bibinfo{volume}{77}}
  , \bibinfo{pages}{235411} (\bibinfo{year}{2008}).

\bibitem[{\citenamefont{G\"u\c{c}l\"u et~al.}(2010)\citenamefont{G\"u\c{c}l\"u, Potasz, and Hawrylak}}]{Guclu+10}
\bibinfo{author}{\bibfnamefont{A.~D.} \bibnamefont{G\"u\c{c}l\"u}},
\bibinfo{author}{\bibfnamefont{P.} \bibnamefont{Potasz}}, \bibnamefont{and}
  \bibinfo{author}{\bibfnamefont{P.}~\bibnamefont{Hawrylak}},
  \bibinfo{journal}{Phys. Rev. B} \textbf{\bibinfo{volume}{82}}
  , \bibinfo{pages}{155445} (\bibinfo{year}{2010}).

\bibitem[{\citenamefont{Nakada et~al.}(1996)\citenamefont{Nakada, Fujita,
  Dresselhaus, and Dresselhaus}}]{NFD+96}
\bibinfo{author}{\bibfnamefont{K.}~\bibnamefont{Nakada}},
  \bibinfo{author}{\bibfnamefont{M.}~\bibnamefont{Fujita}},
  \bibinfo{author}{\bibfnamefont{G.}~\bibnamefont{Dresselhaus}},
  \bibnamefont{and} \bibinfo{author}{\bibfnamefont{M.~S.}
  \bibnamefont{Dresselhaus}}, \bibinfo{journal}{Phys. Rev. B}
  \textbf{\bibinfo{volume}{54}}, \bibinfo{pages}{17954} (\bibinfo{year}{1996}).

\bibitem[{\citenamefont{Fujita et~al.}(1996)\citenamefont{Fujita,Wakabayashi, Nakada, and Kusakabe}}]{Fujita+96}
\bibinfo{author}{\bibfnamefont{M.}~\bibnamefont{Fujita}},
 \bibinfo{author}{\bibfnamefont{K.}~\bibnamefont{Wakabayashi}},
 \bibinfo{author}{\bibfnamefont{K.}~\bibnamefont{Nakada}},
 \bibnamefont{and} \bibinfo{author}{\bibfnamefont{K.}
  \bibnamefont{Kusakabe}}, \bibinfo{journal}{J. Phys. Soc. Jpn.}
  \textbf{\bibinfo{volume}{65}}, \bibinfo{pages}{1920} (\bibinfo{year}{1996}).

\bibitem[{\citenamefont{Son et~al.}(2006)\citenamefont{Son, Cohen, and Louie}}]{Son+06}
\bibinfo{author}{\bibfnamefont{Y.} \bibnamefont{Son}},
  \bibinfo{author}{\bibfnamefont{M.~L.} \bibnamefont{Cohen}}, \bibnamefont{and}
  \bibinfo{author}{\bibfnamefont{S.~G.} \bibnamefont{Louie}},
  \bibinfo{journal}{Nature} \textbf{\bibinfo{volume}{444}},
  \bibinfo{pages}{347} (\bibinfo{year}{2006}).

\bibitem[{\citenamefont{Ezawa}(2006)}]{Ezawa+06}
\bibinfo{author}{\bibfnamefont{M.}~\bibnamefont{Ezawa}},
  \bibinfo{journal}{Phys. Rev. B} \textbf{\bibinfo{volume}{73}}
  , \bibinfo{pages}{045432} (\bibinfo{year}{2006}).

\bibitem[{\citenamefont{Ezawa}(2007)}]{Ezawa+07} 
\bibinfo{author}{\bibfnamefont{M.}~\bibnamefont{Ezawa}},
  \bibinfo{journal}{Phys. Rev. B} \textbf{\bibinfo{volume}{76}}
  , \bibinfo{pages}{245415} (\bibinfo{year}{2007}).

\bibitem[{\citenamefont{Fernandez-Rossier and Palacios}(2007)}]{FRP+07} 
\bibinfo{author}{\bibfnamefont{J.}~\bibnamefont{Fernandez-Rossier}}
  \bibnamefont{and} \bibinfo{author}{\bibfnamefont{J.~J.}
  \bibnamefont{Palacios}}, \bibinfo{journal}{Phys. Rev. Lett.}
  \textbf{\bibinfo{volume}{99}}, \bibinfo{pages}{177204} (\bibinfo{year}{2007}).

\bibitem[{\citenamefont{Akola et~al.}(2008)\citenamefont{Akola, Heiskanen, and
  Manninen}}]{AHM+08}
\bibinfo{author}{\bibfnamefont{J.}~\bibnamefont{Akola}},
  \bibinfo{author}{\bibfnamefont{H.~P.} \bibnamefont{Heiskanen}},
  \bibnamefont{and} \bibinfo{author}{\bibfnamefont{M.}~\bibnamefont{Manninen}},
  \bibinfo{journal}{Phys. Rev. B} \textbf{\bibinfo{volume}{77}}
  , \bibinfo{pages}{193410} (\bibinfo{year}{2008}).

\bibitem[{\citenamefont{Wang et~al.}(2009)\citenamefont{Wang, Yazyev, Meng, and
  Kaxiras}}]{Wang+Yazyev+09}
\bibinfo{author}{\bibfnamefont{W.~L.} \bibnamefont{Wang}},
  \bibinfo{author}{\bibfnamefont{O.~V.}~\bibnamefont{Yazyev}},
  \bibinfo{author}{\bibfnamefont{S.}~\bibnamefont{Meng}}, \bibnamefont{and}
  \bibinfo{author}{\bibfnamefont{E.}~\bibnamefont{Kaxiras}},
  \bibinfo{journal}{Phys. Rev. Lett.} \textbf{\bibinfo{volume}{102}},
  \bibinfo{pages}{157201} (\bibinfo{year}{2009}).

\bibitem[{\citenamefont{Potasz et~al.}(2010)\citenamefont{Potasz, G\"u\c{c}l\"u, and Hawrylak}}]{Potasz+10}
\bibinfo{author}{\bibfnamefont{P.} \bibnamefont{Potasz}}, 
  \bibinfo{author}{\bibfnamefont{A.~D.} \bibnamefont{G\"u\c{c}l\"u}}, \bibnamefont{and}
  \bibinfo{author}{\bibfnamefont{P.}~\bibnamefont{Hawrylak}},
  \bibinfo{journal}{Phys. Rev. B} \textbf{\bibinfo{volume}{81}}
  , \bibinfo{pages}{033403} (\bibinfo{year}{2010}).

\bibitem[{\citenamefont{Niimi et~al.}(2005)\citenamefont{Niimi, Matsui, Kambara, Tagami, Tsukada, and
Fukuyama}}]{Niimi+Matsui+05}
\bibinfo{author}{\bibfnamefont{Y.} \bibnamefont{Niimi}},
  \bibinfo{author}{\bibfnamefont{T.} \bibnamefont{Matsui}},
  \bibinfo{author}{\bibfnamefont{H.} \bibnamefont{Kambara}},
  \bibinfo{author}{\bibfnamefont{K.} \bibnamefont{Tagami}},
  \bibinfo{author}{\bibfnamefont{M.} \bibnamefont{Tsukada}}, \bibnamefont{and}
  \bibinfo{author}{\bibfnamefont{H.} \bibnamefont{Fukuyama}},
  \bibinfo{journal}{Appl. Surf. Sci.} \textbf{\bibinfo{volume}{241}},
  \bibinfo{pages}{43} (\bibinfo{year}{2005}). 

\bibitem[{\citenamefont{Kobayashi et~al.}(2005)\citenamefont{Kobayashi, Fukui, Enoki, Kusakabe, and Kaburagi}}]{Kobayashi+Fukui+05}
\bibinfo{author}{\bibfnamefont{Y.}~\bibnamefont{Kobayashi}},
  \bibinfo{author}{\bibfnamefont{K.-I.}~\bibnamefont{Fukui}},
  \bibinfo{author}{\bibfnamefont{T.}~\bibnamefont{Enoki}},
  \bibinfo{author}{\bibfnamefont{K.}~\bibnamefont{Kusakabe}}, \bibnamefont{and}
  \bibinfo{author}{\bibfnamefont{Y.}~\bibnamefont{Kaburagi}},
  \bibinfo{journal}{Phys. Rev. B} \textbf{\bibinfo{volume}{71}},
  \bibinfo{pages}{193406} (\bibinfo{year}{2005}).

\bibitem{Tao+Jiao+11}
C. Tao, L. Jiao, O.V. Yazyev, Y.-C. Chen, J. Feng, X. Zhang, R.B. Capaz, J.M. Tour, A. Zettl, S.G. Louie, H. Dai, and M.F. Crommie,
{Nature Phys.} {\bf 7}, 616 (2011).

\bibitem[{\citenamefont{Wang et~al.}(2008)\citenamefont{Wang, Meng, and
  Kaxiras}}]{Wang+Meng+08}
\bibinfo{author}{\bibfnamefont{W.~L.} \bibnamefont{Wang}},
  \bibinfo{author}{\bibfnamefont{S.}~\bibnamefont{Meng}}, \bibnamefont{and}
  \bibinfo{author}{\bibfnamefont{E.}~\bibnamefont{Kaxiras}},
  \bibinfo{journal}{Nano Letters} \textbf{\bibinfo{volume}{8}},
  \bibinfo{pages}{241} (\bibinfo{year}{2008}).

\bibitem[{\citenamefont{G\"u\c{c}l\"u et~al.}(2009)\citenamefont{G\"u\c{c}l\"u, Potasz, Voznyy, Korkusinski, and
  Hawrylak}}]{Guclu+09}
\bibinfo{author}{\bibfnamefont{A.~D.} \bibnamefont{G\"u\c{c}l\"u}},
  \bibinfo{author}{\bibfnamefont{P.} \bibnamefont{Potasz}},
  \bibinfo{author}{\bibfnamefont{O.}~\bibnamefont{Voznyy}},
  \bibinfo{author}{\bibfnamefont{M.} \bibnamefont{Korkusinski}}, \bibnamefont{and}
  \bibinfo{author}{\bibfnamefont{P.}~\bibnamefont{Hawrylak}},
  \bibinfo{journal}{Phys. Rev. Lett.} \textbf{\bibinfo{volume}{103}}
  , \bibinfo{pages}{246805} (\bibinfo{year}{2009}).

\bibitem[{\citenamefont{Ezawa}(2008)}]{Ezawa+08}
\bibinfo{author}{\bibfnamefont{M.}~\bibnamefont{Ezawa}},
  \bibinfo{journal}{Phys. Rev. B} \textbf{\bibinfo{volume}{77}}
  , \bibinfo{pages}{155411} (\bibinfo{year}{2008}). 

\bibitem[{\citenamefont{Philpott et~al.}(2008)\citenamefont{Philpott, Cimpoesu, and Kawazoe}}]{Philpott+08}
\bibinfo{author}{\bibfnamefont{M.~R.} \bibnamefont{Philpott}},
  \bibinfo{author}{\bibfnamefont{F.}~\bibnamefont{Cimpoesu}}, \bibnamefont{and}
  \bibinfo{author}{\bibfnamefont{Y.}~\bibnamefont{Kawazoe}},
  \bibinfo{journal}{Chem. Phys.} \textbf{\bibinfo{volume}{354}},
  \bibinfo{pages}{1} (\bibinfo{year}{2008}).

\bibitem[{\citenamefont{Heiskanen et~al.}(2008)\citenamefont{Heiskanen, Manninen, and Akola}}]{HMA+08}
\bibinfo{author}{\bibfnamefont{H.~P.} \bibnamefont{Heiskanen}},
 \bibnamefont{and} \bibinfo{author}{\bibfnamefont{M.}~\bibnamefont{Manninen}},
 \bibinfo{author}{\bibfnamefont{J.}~\bibnamefont{Akola}},
  \bibinfo{journal}{New J. Phys.} \textbf{\bibinfo{volume}{10}}
  , \bibinfo{pages}{103015} (\bibinfo{year}{2008}).

\bibitem[{\citenamefont{Kosimov et~al.}(2010)\citenamefont{Kosimov, Dzhurakhalov, and Peeters}}]{Kosimov+10}
\bibinfo{author}{\bibfnamefont{D.~P.} \bibnamefont{Kosimov}}, 
  \bibinfo{author}{\bibfnamefont{A.~A.} \bibnamefont{Dzhurakhalov}}, \bibnamefont{and}
  \bibinfo{author}{\bibfnamefont{F.~M.}~\bibnamefont{Peeters}},
  \bibinfo{journal}{Phys. Rev. B} \textbf{\bibinfo{volume}{81}}
  , \bibinfo{pages}{195414} (\bibinfo{year}{2010}).

\bibitem[{\citenamefont{Ezawa}(2010)}]{Ezawa+10}
\bibinfo{author}{\bibfnamefont{M.}~\bibnamefont{Ezawa}},
  \bibinfo{journal}{Phys. Rev. B} \textbf{\bibinfo{volume}{81}}
  , \bibinfo{pages}{201402} (\bibinfo{year}{2010}).

\bibitem[{\citenamefont{Ezawa}(2010)}]{Ezawa+E10}
\bibinfo{author}{\bibfnamefont{M.}~\bibnamefont{Ezawa}},
  \bibinfo{journal}{Physica E} \textbf{\bibinfo{volume}{42}}
  , \bibinfo{pages}{703} (\bibinfo{year}{2010}).

\bibitem[{\citenamefont{Voznyy et~al.}(2011)\citenamefont{Voznyy, G\"u\c{c}l\"u, Potasz, and
  Hawrylak}}]{Voznyy+11}
  \bibinfo{author}{\bibfnamefont{O.}~\bibnamefont{Voznyy}},
\bibinfo{author}{\bibfnamefont{A.~D.} \bibnamefont{G\"u\c{c}l\"u}},
  \bibinfo{author}{\bibfnamefont{P.} \bibnamefont{Potasz}},
 \bibnamefont{and}
  \bibinfo{author}{\bibfnamefont{P.}~\bibnamefont{Hawrylak}},
  \bibinfo{journal}{Phys. Rev. B} \textbf{\bibinfo{volume}{83}}
  , \bibinfo{pages}{165417} (\bibinfo{year}{2011}).

\bibitem[{\citenamefont{Sahin et~al.}(2010)\citenamefont{Sahin, Senger, and Ciraci}}]{Sahin+10}
\bibinfo{author}{\bibfnamefont{H.} \bibnamefont{Sahin}}, 
  \bibinfo{author}{\bibfnamefont{R.~T.} \bibnamefont{Senger}}, \bibnamefont{and}
  \bibinfo{author}{\bibfnamefont{S.}~\bibnamefont{Ciraci}},
  \bibinfo{journal}{J. Appl. Phys.} \textbf{\bibinfo{volume}{108}}
  , \bibinfo{pages}{074301} (\bibinfo{year}{2010}).

\bibitem[{\citenamefont{Romanovsky et~al.}(2011)\citenamefont{Romanovsky, Yannouleas, and Landman}}]{Romanovsky+11}
\bibinfo{author}{\bibfnamefont{I.} \bibnamefont{Romanovsky}}, 
  \bibinfo{author}{\bibfnamefont{C.} \bibnamefont{Yannouleas}}, \bibnamefont{and}
  \bibinfo{author}{\bibfnamefont{U.}~\bibnamefont{Landman}},
  \bibinfo{journal}{Phys. Rev. B} \textbf{\bibinfo{volume}{83}}
  , \bibinfo{pages}{045421} (\bibinfo{year}{2011}).

\bibitem{Xi+09}
Y. Xi, M. Zhao, X. Wang, S. Li, X. He, Z. Wang, and H. Bu,
{J. Phys. Chem. C} {\bf 113}, 12637 (2009).

\bibitem{Kinza+10}
M. Kinza, J. Ortloff, and C. Honerkamp,
{Phys. Rev. B} {\bf 82}, 155430 (2010).

\bibitem{Zarenia+11}
M. Zarenia, A. Chaves, G. A. Farias, and F. M. Peeters,
{Phys. Rev. B} {\bf 84}, 245403 (2011).

\bibitem{Dai+12}
Q. Q. Dai, Y. F. Zhu and Q. Jiang,
{Phys. Chem. Chem. Phys.} {\bf 14}, 1253–1261 (2012).

  \bibitem[{\citenamefont{Lieb}(1989)\citenamefont{Lieb}}]{Lieb+89}
\bibinfo{author}{\bibfnamefont{E.~H.}~\bibnamefont{Lieb}},
  \bibinfo{journal}{Phys. Rev. Lett.} \textbf{\bibinfo{volume}{62}},
  \bibinfo{pages}{1201} (\bibinfo{year}{1989}).

\bibitem[{\citenamefont{Wallace}(1947)}]{Wallace+47}
\bibinfo{author}{\bibfnamefont{P.~R.}~\bibnamefont{Wallace}},
  \bibinfo{journal}{Phys. Rev.} \textbf{\bibinfo{volume}{71}}, \bibinfo{pages}{622} (\bibinfo{year}{1947}).

\bibitem[{\citenamefont{Ransil}(1960)}]{Ransil+60}
\bibinfo{author}{\bibfnamefont{B.~J.}~\bibnamefont{Ransil}},
  \bibinfo{journal}{Rev. Mod. Phys.} \textbf{\bibinfo{volume}{32}}, \bibinfo{pages}{245} (\bibinfo{year}{1960}).

\bibitem[{\citenamefont{Potasz et~al.}(2010)\citenamefont{Potasz, G\"u\c{c}l\"u, and Hawrylak}}]{Potasz+ring+10}
\bibinfo{author}{\bibfnamefont{P.} \bibnamefont{Potasz}}, 
  \bibinfo{author}{\bibfnamefont{A.~D.} \bibnamefont{G\"u\c{c}l\"u}}, \bibnamefont{and}
  \bibinfo{author}{\bibfnamefont{P.}~\bibnamefont{Hawrylak}},
  \bibinfo{journal}{Phys. Rev. B} \textbf{\bibinfo{volume}{82}}
  , \bibinfo{pages}{075425} (\bibinfo{year}{2010}).

\bibitem[{\citenamefont{Reich et~al.}(2002)\citenamefont{Reich, Maultzsch, Thomsen, and Ordej\'{o}n}}]{Reich+02}
\bibinfo{author}{\bibfnamefont{S.}~\bibnamefont{Reich}},
\bibinfo{author}{\bibfnamefont{J.}~\bibnamefont{Maultzsch}},
\bibnamefont{and}
\bibinfo{author}{\bibfnamefont{C.}~\bibnamefont{Thomsen}},
\bibnamefont{and}
\bibinfo{author}{\bibfnamefont{P.}~\bibnamefont{Ordej\'{o}n}},
  \bibinfo{journal}{Phys. Rev. B} \textbf{\bibinfo{volume}{66}}, \bibinfo{pages}{035412} (\bibinfo{year}{2002}).

\bibitem[{\citenamefont{Bostwick et~al.}(2007)\citenamefont{Bostwick, Ohta, McChesney, Seyller, Horn, and Rotenberg}}]{Bostwick+07}
\bibinfo{author}{\bibfnamefont{A.}~\bibnamefont{Bostwick}},
\bibinfo{author}{\bibfnamefont{T.}~\bibnamefont{Ohta}},
\bibinfo{author}{\bibfnamefont{J.~L.}~\bibnamefont{McChesney}},
\bibinfo{author}{\bibfnamefont{T.}~\bibnamefont{Seyller}},
\bibinfo{author}{\bibfnamefont{K.}~\bibnamefont{Horn}},
\bibnamefont{and}
\bibinfo{author}{\bibfnamefont{E.}~\bibnamefont{Rotenberg}},
  \bibinfo{journal}{Solid State Commun.} \textbf{\bibinfo{volume}{143}}, \bibinfo{pages}{63} (\bibinfo{year}{2007}).

\bibitem[{\citenamefont{Deacon et~al.}(2007)\citenamefont{Deacon, Chuang, 
Nicholas, Novoselov, and Geim}}]{DCN+07}
\bibinfo{author}{\bibfnamefont{R.~S.} \bibnamefont{Deacon}},
  \bibinfo{author}{\bibfnamefont{K.-C.} \bibnamefont{Chuang}},
  \bibinfo{author}{\bibfnamefont{R.J.}~\bibnamefont{Nicholas}},
  \bibinfo{author}{\bibfnamefont{K.~S.}~\bibnamefont{Novoselov}}, \bibnamefont{and}
  \bibinfo{author}{\bibfnamefont{A.K.}~\bibnamefont{Geim}},
  \bibinfo{journal}{Phys. Rev. B} \textbf{\bibinfo{volume}{76}}, 
  \bibinfo{pages}{081406(R)} (\bibinfo{year}{2007}).

\bibitem[{\citenamefont{Wunsch et~al.}(2008)}]{Wunsch+2008} 
\bibinfo{author}{\bibfnamefont{B.}~\bibnamefont{Wunsch}}
\bibinfo{author}{\bibfnamefont{T.}~\bibnamefont{Stauber}},
\bibnamefont{and}
\bibinfo{author}{\bibfnamefont{F.}~\bibnamefont{Guinea}}, 
\bibinfo{journal}{Phys. Rev. B} \textbf{\bibinfo{volume}{77}}, \bibinfo{pages}{035316} (\bibinfo{year}{2008}).

\bibitem[{\citenamefont{Romanovsky et~al.}(2009)}]{Romanovsky+2009} 
\bibinfo{author}{\bibfnamefont{I.}~\bibnamefont{Romanovsky}}
\bibinfo{author}{\bibfnamefont{Y.}~\bibnamefont{Yannouleas}},
\bibnamefont{and}
\bibinfo{author}{\bibfnamefont{U.}~\bibnamefont{Landman}}, 
\bibinfo{journal}{Phys. Rev. B} \textbf{\bibinfo{volume}{79}}, \bibinfo{pages}{075311} (\bibinfo{year}{2009}).

\bibitem[{\citenamefont{Hawrylak et~al.}(1996)}]{Hawrylak+Wojs+Brum+96} 
\bibinfo{author}{\bibfnamefont{P.}~\bibnamefont{Hawrylak}}
\bibinfo{author}{\bibfnamefont{A.}~\bibnamefont{Wojs}},
\bibnamefont{and}
\bibinfo{author}{\bibfnamefont{J.~A.}~\bibnamefont{Brum}}, 
\bibinfo{journal}{Phys. Rev. B} \textbf{\bibinfo{volume}{55}}, \bibinfo{pages}{11397} (\bibinfo{year}{1996}).

\end{thebibliography}

\end{document}